\documentclass[12pt]{iopart}

\usepackage{color,colordvi,graphicx}

\begin{document}

\title[Para/ortho-H$_2$+D reactions]{Effect of nuclear spin symmetry
  in cold and ultracold reactions: D + para/ortho-H$_2$}

\author{Ionel Simbotin and Robin C\^ot\'e}

\address{Department of Physics, University of Connecticut, 2152
  Hillside Rd., Storrs, CT 06269, USA} \ead{rcote@phys.uconn.edu}
\vspace{10pt}
\begin{indented}
\item[]January 2015
\end{indented}

\begin{abstract}
  We report results for reaction and vibrational quenching of the
  collision D with para-H$_2$($v,j=0$) and ortho-H$_2$($v,j=1$) at
  cold and ultracold temperatures. We investigate the effect of
  nuclear spin symmetry for barrier dominated processes ($0\le v\le
  4$) and for one barrierless case ($v=5$). We find resonant
  structures for energies in the range corresponding to
  0.01--10 K, which depend on the nuclear spin of H$_2$,
  arising from contributions of specific partial waves.  We discuss
  the implications on the results in this benchmark system for
  ultracold chemistry.
\end{abstract}

%
%
%
%
%

\section{Introduction}

Research with cold and ultracold molecules has witnessed an explosive
growth since the first predictions \cite{cote-1997,jmp-1999} and
observations \cite{pillet-1998,knize-1998} of ultracold molecules in
the late 1990s. Already, several
books~\cite{gospel:09,ian.smith:08,w+z:09} have been published on cold
and ultracold molecules, and numerous reviews have been written on
more specialized topics
\cite{stwalley:canjchem:04,weck+bala:irpc:06,jeremy:irpc:06,
  jeremy:irpc:07,krems:pccp:08}, illustrating the continuously
broadening scope of this field, as exemplified by this special issue.

In addition to new proposals to produce ultracold molecules (as in
Ref.~\cite{FOPA+STIRAP}), this rapidly expanding field has motivated
theoretical studies of the chemical structure and properties of
ultracold molecules, {\it e.g.}, for the interaction of two KRb
molecules \cite{Thermo-K2Rb2}.  The extraordinary level of control of
all degrees of freedom, from internal to translational degrees, reached in
ultracold systems has allowed the investigation of the effect of the
nuclear spin on atom-diatom scattering, as demonstrated in
Ref.~\cite{junye:sci:10} with KRb+K. 

In this article, we explore such questions using a benchmark system
exhibiting both quenching and reaction, namely H$_2$+D, which can
react to produce HD+H. Because of its small mass, this system offers
the opportunity to study the contributions of low partial waves even
at cold temperatures, as opposed to heavier systems where ultracold
temperatures are required.  This particular system is fundamental in
quantum chemistry, and is being actively investigated in experiments
capable of reaching the energy range explored in this work
\cite{zare-private}.  The detection of the resonance features we
predict would be a valuable test of the {\it ab initio} potential
energy surface calculation, and their dependence on the nuclear
symmetry provide an even stricter comparison between experiment and
theory, the need of which is discussed in Ref.\cite{zare-PNAS}.  In
addition, H$_2$+D is also relevant to astrophysics, especially in the
astrochemistry of cold interstellar clouds \cite{astro-cloud}, and
even in the early universe \cite{astro}. Although many studies
involving the scattering of H$_2$ with various atoms have been
performed in the ultracold regime
\cite{Schatz:JCP:1987,bala:fh2:cpl01,bodo:fd2:jpb02,Bala:PRL:Heh2,bala:KrArH2:pra09},
the effect of the nuclear symmetry has not been studied for this
reactive system at low temperatures.  However, nuclear symmetry has
been considered in studies of para- and ortho-H$_2$, e.g., the
rovibrational relaxation and energy transfer in ultracold H$_2$+H$_2$
collisions \cite{bala-PRA2008,bala-JCP2009,bala-JCP2011}, and in
studies of vibrational quenching of O$_2$ when colliding with He
\cite{bohn-PRA2001}, or spin-changing scattering between two ground
state O$_2$ molecules \cite{bohn-JCP2003}.

In Sec.~\ref{sec:theory} we give a brief description of the
theoretical and numerical tools used, as well as the properties of
this benchmark system. We present and analyze the results in
Sec.~\ref{sec:results}, and conclude in Sec.~\ref{sec:conclusion}.

\section{Theoretical and computational details}
\label{sec:theory}

We are interested in the effect of the nuclear spin symmetry of
homonuclear molecules in reactive scattering processes at low
temperatures. To this end, we consider the benchmark reaction H$_2$+D
which involves the simplest and most fundamental diatomic molecule,
H$_2$.  The nuclear part of the wave function of H$_2$ must be
anti-symmetric with respect to the permutation of the two protons, and
since the nuclear spin wave function of two spin $i=1/2$ nuclei with
total nuclear spin $I$ acquires a factor $(-1)^{2i-I}$ under exchange,
the rotational states of molecular hydrogen are restricted:
para-Hydrogen corresponds to the singlet spin state with ($I=0,
M_I=0$) which allows only even-$j$ rotational states, while the
nuclear triplet spin state of ortho-Hydrogen with ($I=1, M_I=0,\pm 1$)
allows only odd-$j$ rotational states.

\subsection{Numerical approach}
\label{sec:numerical}

The expression for the state-to-state cross sections, integrated over
all scattering directions, averaged over the initial rotational states
of the reactant dimer, and summed over the final rotational states of
the product, reads
\begin{equation}
\label{eq:sigma}
\sigma_{n'\leftarrow n}(E) = \frac{\pi}{k^{2}_n}
	\sum_{J=0}^{\infty} \left(\frac{2J+1}{2j+1}\right)
	\sum_{\ell = |J-j|}^{|J+j|}
    \;\;\sum_{\ell'= |J-j'|}^{|J+j'|}
     \left| T^{J}_{n'\ell'  n\ell}(E)\right|^{2},
\end{equation}
where the generic notation $n=({\rm a} v j)$ stands for the
arrangement label ``a" and quantum numbers $(vj)$ of the molecular
H$_2$ states, and $k_n=\sqrt{2\mu_{\rm a} E_{\rm kin}}$ is the initial
momentum ($\hbar=1$, atomic units are used); $\mu_a$ the reduced mass
of the binary system atom--diatom in the initial arrangement (a), the
initial kinetic energy is $E_{\rm kin}=E-\varepsilon_n$, and
$\varepsilon_n$ are the rovibrational energies of the dimers.  $E$ is
the (total) collision energy, $J$ is the total angular momentum, and
$\ell$ is angular momentum for the relative motion.  The primes
indicate the corresponding quantities in the exit channel, with
$n'=({\rm a}' v' j')$ and angular momentum $\ell'$.  The T-matrix
$T^{J}_{n'\ell'\;n\ell}=\delta_{n'n}\delta_{\ell'\ell} -
S^{J}_{n'\ell'\; n\ell}$ is given in terms of the $S$-matrix.

We obtain an energy dependent rate constant by multiplying the cross
sections by the relative velocity $v_{\rm rel.}$ in the initial
arrangement $a$ (D + para/ortho-H$_2$) of the incoming scattering
channel, namely
\begin{equation}
\label{eq:rate}
{\sf K}_{n'\leftarrow n}(E) = v_{\rm rel.} \sigma_{n'\leftarrow n}(E) \; ,
\end{equation}
where $v_{\rm rel.}= \sqrt{2E_{\rm kin}/\mu_{\rm a}}$. The total rate
constant is obtained by summing over the appropriate final states
$n'$. If the arrangement remains the same ($a'=a$), the scattering
process corresponds to quenching of H$_2$ by D to lower states, and if
$a'\neq a$, a reaction occurred leading to HD + H. The corresponding
rate constants are simply
\begin{eqnarray}
 {\sf K}_n^{\rm Q}(E) & = & \sum_{n'\neq n, a'=a} {\sf K}_{n'\leftarrow n}(E) \;, \label{eq:K-Q} \\
 {\sf K}_n^{\rm R}(E) & = & \sum_{n'\neq n, a'\neq a} {\sf K}_{n'\leftarrow n}(E) \label{eq:K-R} \;.
\end{eqnarray}
We note that if $n'=n$, we obtain an elastic collision: this is
discussed in more detail in \S~\ref{sec:elastic}.

\begin{figure}[t]
\includegraphics[clip,width= 1.0 \textwidth]{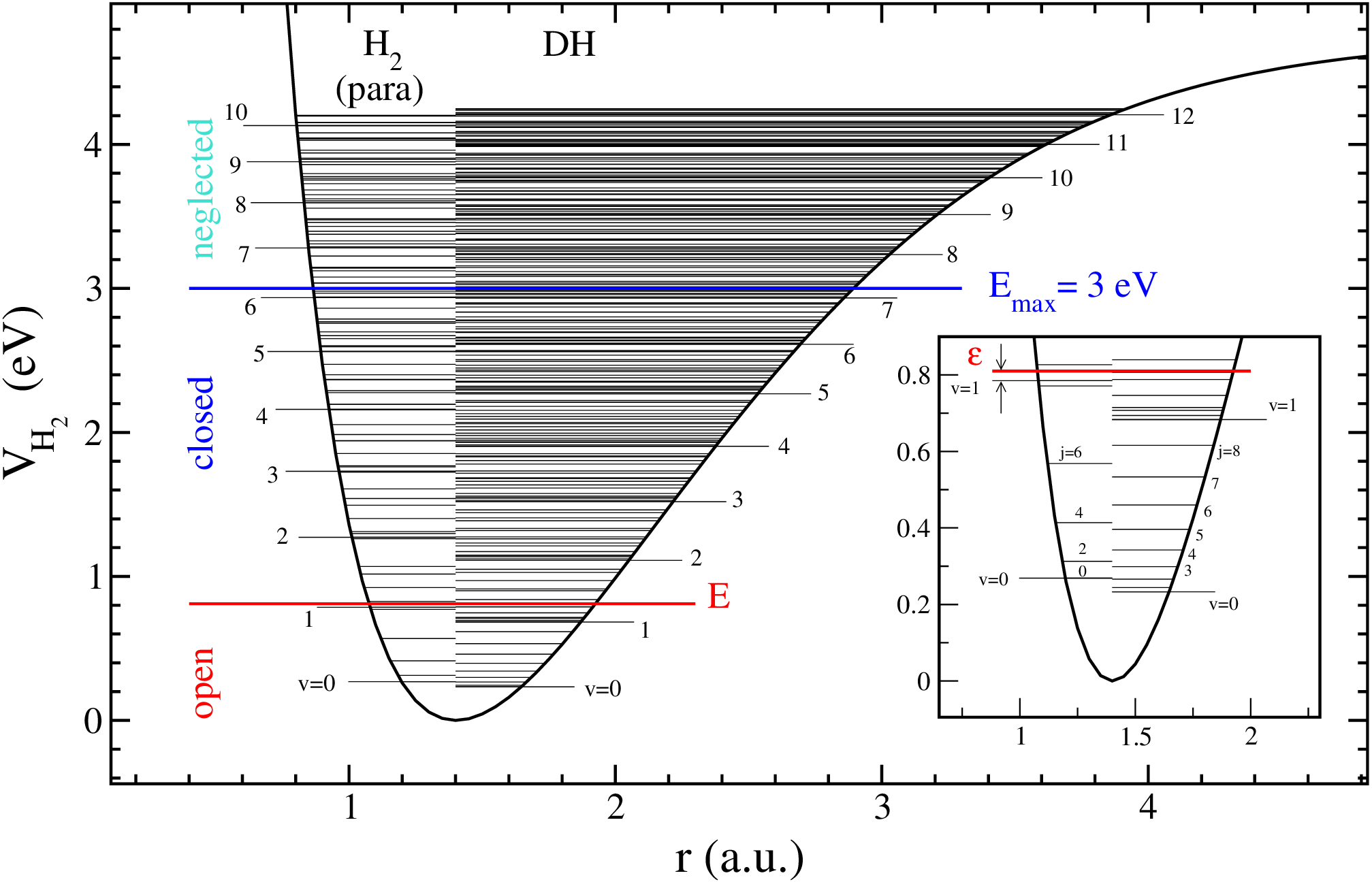}
\caption{Ro-vibrational levels of para-H$_2$ (left),
  and HD (right). Relevant channels for H$_2$ initially in $v=1$ are
  shown: open channels with energies smaller than the absolute
  scattering energy $E$ for quenching and reactions, closed channels
  with energies larger than $E$ affecting the results, and the
  neglected channels above a certain truncation energy $E_{\max}$.
  The inset shows a zoom of the levels near $v=1$, with the scattering
  energy $\varepsilon$ relative to the entrance channel $(v=1,j=0)$. }
\label{fig:Levels}
\end{figure}

To solve for the $S$-matrix and thus Eq.(\ref{eq:sigma}), we modified
the ABC reactive scattering code developed by Manolopoulos and
coworkers \cite{abc:cpc:2k}, which is based on Delves \cite{Delves:58}
hyperspherical coordinates, with a separate set of Delves
hyperspherical coordinates for each arrangement sharing the same
hyperradius coordinate.  The full wave function is expanded in
eigenbases corresponding to internal motion coordinates for each
arrangement (after the combined multiple-arrangement basis is
orthogonalized), and the resulting hyperradial coupled-channel
equations are solved using the log derivative method
\cite{manolopoulos:logder:jcp86}. The propagation starts at
$\rho=\rho_{\min}$ (where the potential is highly repulsive), and
ends at a sufficiently large $\rho=\rho_{\max}$, where the S-matrix
is extracted by imposing asymptotic boundary conditions. We note that
the ABC code is well suited for system involving barriers; it was
tested at high energies for a number of benchmark systems such as H +
H$_2$, F + H$_2$, Cl + H$_2$, and their isotopic counterparts
\cite{miranda:hd2:jcp98,jesus:hd2:pccp10,castillo:fh2:jcp96,
  mano:fhd:fdiscuss98,skouteris:clhd:science99}, and modified versions
to study certain benchmark chemical reactions in the ultracold regime
\cite{bala:fh2:cpl01,bala:clhd:JCP04,bodo:fd2:jpb02,
  bala:fhcl:jcp08,jesus:fh2:jcp06}.

\begin{figure}[t]
\includegraphics[clip,width=0.54\textwidth]{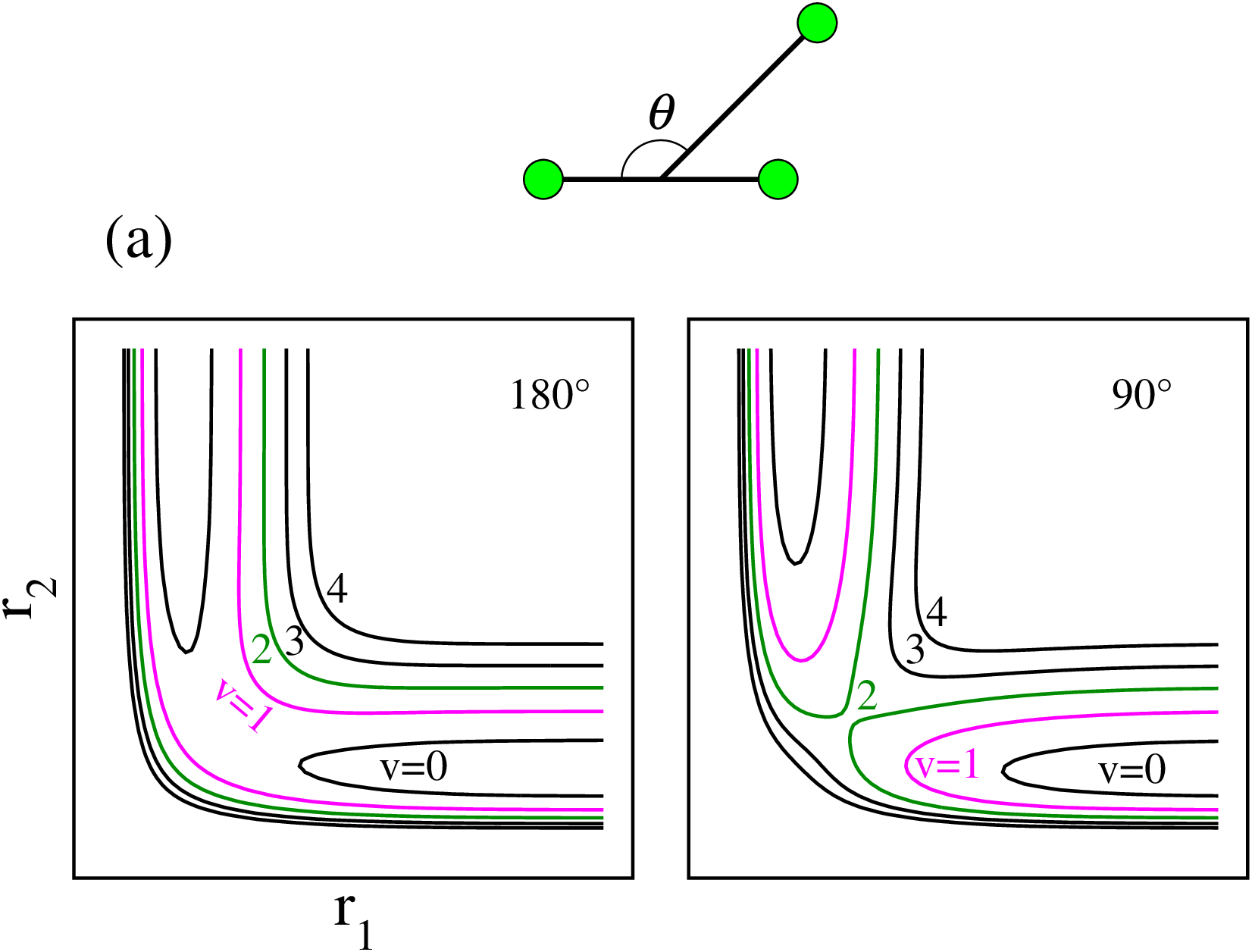}
\hfill
\includegraphics[clip,width=0.44\textwidth]{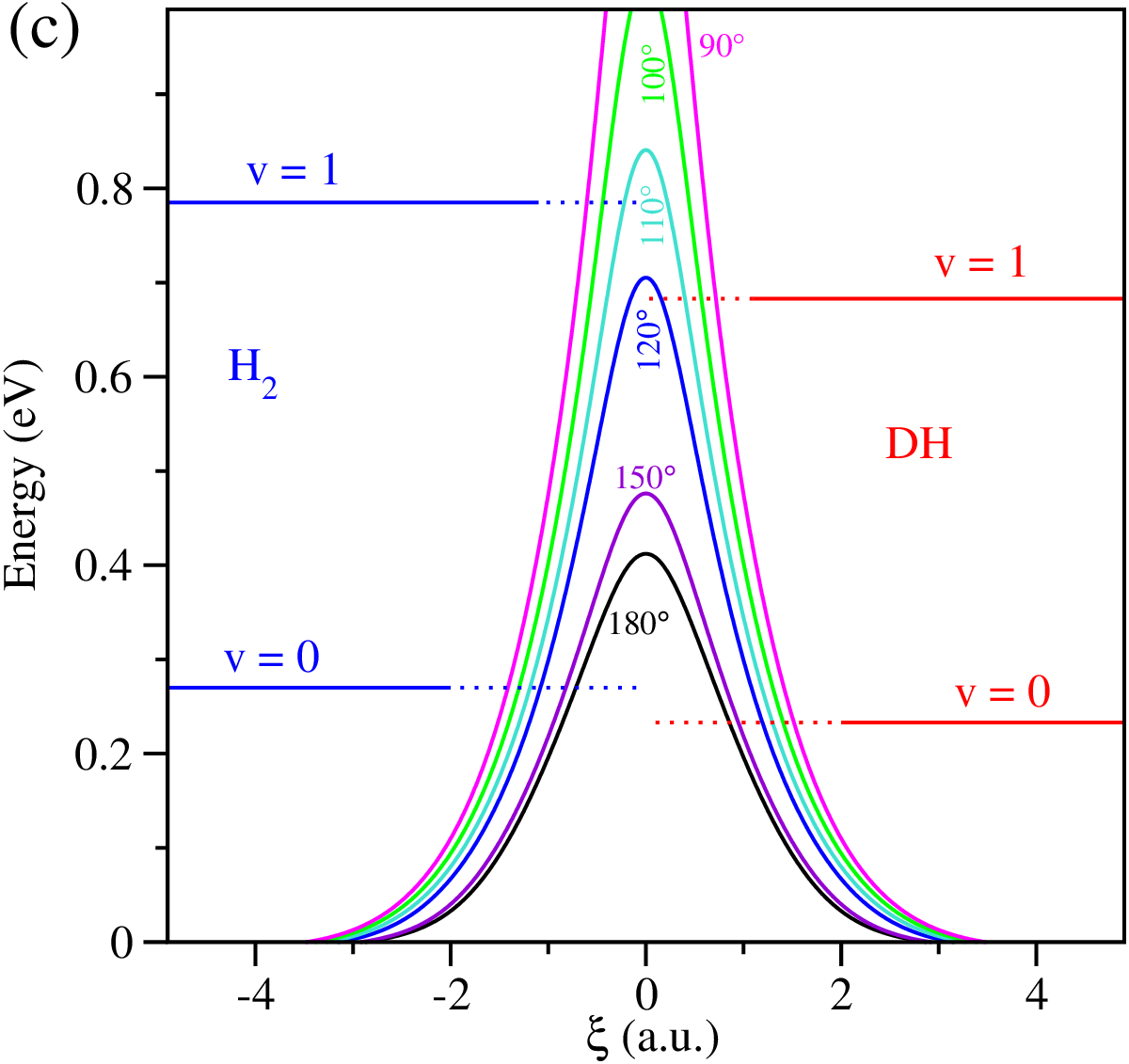}
\includegraphics[clip,width=\textwidth]{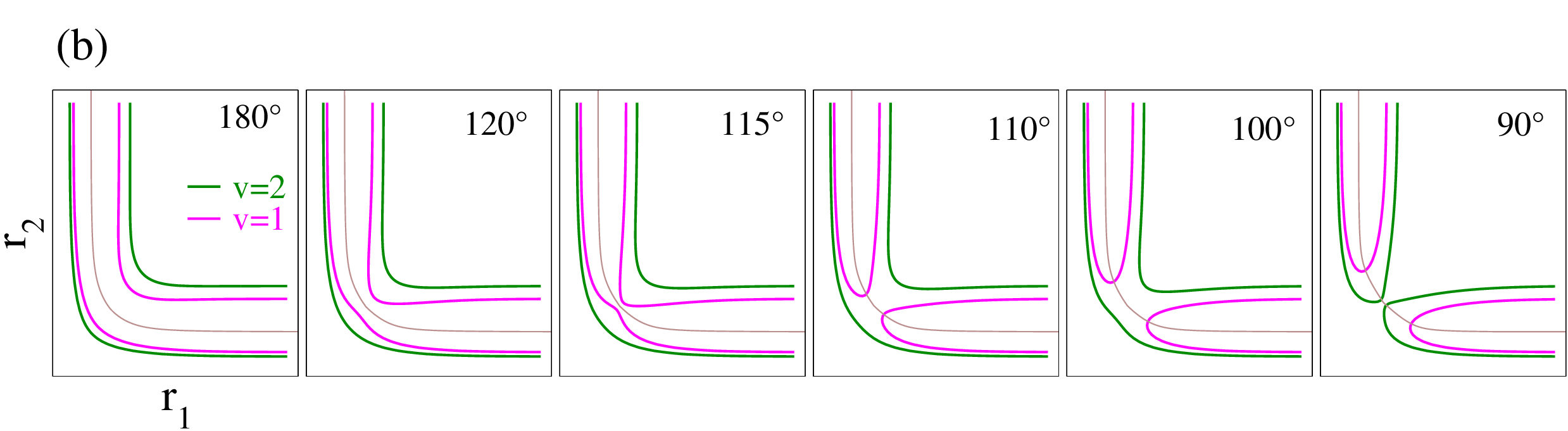}
\caption{(a) Contour plot of the PES for H$_3$ with the
  vibrational levels $v=0, \cdots, 4$ plotted for two geometries
  ($\theta = 180^{\rm o}$ and $90^{\rm o}$ as defined in the sketch
  above). This plot shows that the barrier is always present for
  $v=0$, and partially for $v=1$ and 2. For higher $v$ its influence
  is reduced.  (b) Progression of the barrier as $\theta$ varies: it
  appears between $110^{\rm o}$ and $115^{\rm o}$ for $v=1$, and only
  near $90^{\rm o}$ for $v=2$. The line passing through the middle
  shows the reaction coordinate $\xi$.  (c) Energy surface along the
  reaction coordinate $\xi$ for different angle $\theta$.  The
  position of the vibrational levels $v=0$ and 1 (with $j=0$) for
  H$_2$ and HD are also shown; $v=0$ of both is always below the top
  of the barrier, indicating that reactions require tunneling, while
  $v=1$ of H$_2$ is above the barrier for $\theta > 112^{\rm o}$. }
\label{fig:contour-combined}
\end{figure}

To obtain fully converged results is expensive computationally, and a
sufficient number of terms in the sum in Eq.(\ref{eq:sigma}) are
required for higher energies.  In Sec.~3 our results will show clearly
that a maximum value of $J_{\max}=4$ is required for energies
in the kelvin regime, i.e., for $E\leq 10$~K.  For much higher
energies, which are outside the scope of this work, the value of
$J_{\max}$ increases steadily.  We remark that although the
blocks for each $J$ are separate (due to conservation of total angular
momentum), the coupled problem for a given $J$ becomes rapidly
prohibitive as $J$ increases.  As described in \cite{PCCP-H2+D}, we
modified the ABC code's structure and adapted it to the cold and
ultracold regime; most important in our implementation is the ability
to follow the progress of convergence in great detail with a
save-and-restart feature, which allows us to go back and extend the
radial propagation of a given run if needed.  We typically monitor the
convergence simultaneously for a large number of collision energies
and for many initial states in a single run. Typically, we propagate
to $\rho_{\max} \ge$ 40~a.u. to obtain converged results using
small integration step size ($\Delta\rho$ $\le$ 0.002 a.u.)
\cite{PCCP-H2+D}.  Finally, in addition to truncating the sum over $J$
(and partial waves), we also restrict the number of channels by fixing
a maximum energy $E_{\max}$.  As illustrated in
Fig.~\ref{fig:Levels} for the case of the ($v=1,j=0$) initial channel
of H$_2$, closed channels above $E_{\max}$ are neglected. This
figure also depicts the difference in the number of molecular states
between para-H$_2$ and HD; The density of states in HD is larger not
only because it is more massive, but also because the restriction on
rotational states $j$ is lifted.

\subsection{Interactions and Potential Energy Surface (PES)}
\label{sec:PES}

The benchmark system we consider here, H$_2$+D, has already been
studied at ultracold \cite{PCCP-H2+D} and higher temperatures
\cite{mielke}.  Accurate {\it ab initio} potential energy
surfaces (PES) exist \cite{bkmp2:jcp96,mielke} for this system. As in
our previous studies \cite{PCCP-H2+D,PRL-rydberg-dressing}, we adopted
the electronic ground state PES of Ref.\cite{bkmp2:jcp96}.

This system possesses an activation barrier, which is illustrated by a
contour plot of the vibrational levels in
Fig.~\ref{fig:contour-combined}(a,b); basically, the barrier is
present for any geometry for $v=0$, requiring tunneling for reactions
to occur, while $v=1$ is partially affected by the barrier ({\it i.e.}
the barrier is present for a finite range of angles only). For higher
$v$, the barrier does not block reactions (although it still affects
the scattering rates). This figure also shows the reaction coordinates
$\xi$. Following the line $\xi$, we plot the barrier for different
angles $\theta$ and the lowest energy levels of H$_2$ and HD in
Fig.~\ref{fig:contour-combined}(c). We can clearly see that $v=0$
requires tunneling to produce HD, while $v=1$ is higher than the
barrier at angles starting at $112^{\rm o}$.

Hyperfine interactions are not considered in our treatment.  Several
studies have pointed out the effect of hyperfine coupling in cold and
ultracold molecular dynamics, mainly in the inelastic processes of
non-reactive systems such as YbF+He \cite{krems-PRA-2007}, NH+Mg
\cite{Hutson-PRA-2011}, MnH+He \cite{Halvick-PCCP-2011}, or in
reactive systems such as Na$_2$+Na in which only the atomic hyperfine
state is modified while the molecular hyperfine state remains
unchanged \cite{Launay-Laser-2006}. Other studies involving H$_2$ at
higher temperatures for the interpretation of astrophysical spectra,
such as CN or HCN interacting with para-H$_2$ \cite{Lique-Astro-2012},
or HCl with para- or ortho-H$_2$ \cite{Lique-JCP-2014}, give
approximate methods to account for hyperfine interactions. However,
like in our case, the hyperfine structure of H$_2$ is not
considered. The H$_2$ hyperfine splittings in its electronic ground
state X$^1\Sigma_g^+$ are negligible, so that the H$_2$+D scattering
will be dominated by the hyperfine structure of D (327 MHz or 15.7
mK). Since the focus of this work is the structure found in the range
of 100 mK to roughly 10 K, the effect of hyperfine interaction can
safely be neglected in our study of nuclear symmetry effects.
However, at sub-millikelvin temperatures, the hyperfine effects will
become relevant, in particular if a resonance is located in the regime
of extremely low energies.

\section{Results and discussion}
\label{sec:results}

We give here an overview of the influence of the nuclear spin state on
the scattering properties (reaction, quenching, and elastic processes)
of H$_2$ with D. The nuclear singlet spin state of para-Hydrogen with
($I=0, M_I=0$) allows only even-$j$ rotational states, while the
nuclear triplet spin state of ortho-Hydrogen with ($I=1, M_I=0,\pm 1$)
allows only odd-$j$ rotational states. We consider here only the
lowest rotational states $j=0$ and $j=1$ for the initial para- and
ortho-H$_2$, respectively. In what follows, we show results for
scattering energies ranging from 1 $\mu$K to 100 K; as mentioned in
the previous section, hyperfine interactions will become relevant for
energies below roughly 15 mK, and thus our results below that energy
are approximate.

\subsection{Overview}
\label{sec:overview}

\begin{figure}[t]
\includegraphics[clip,width= 1.0 \textwidth]{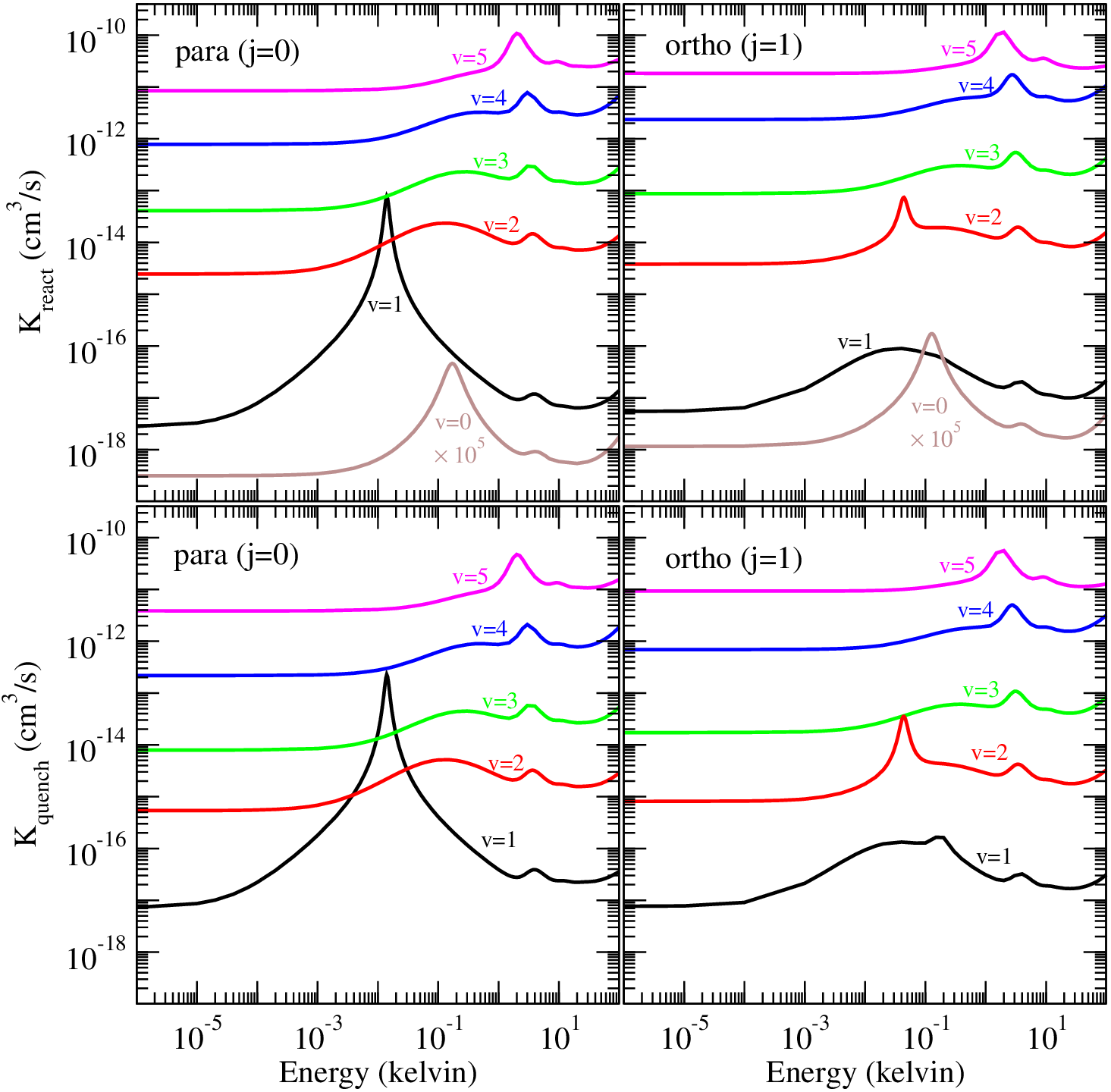}
\caption{Rate constant ${\sf K}_v(E)$ ($v\leq 5$)
  vs. collision energy corresponding to the range of 1 $\mu$K to 100 K
  for reaction (top panels) and quenching (bottom panels) processes
  for para-H$_2$ (left panels) and ortho-H$_2$ (right panels)
  interacting with D.}
\label{fig:overvu-4panel}
\end{figure}

We computed the energy-dependent rate constant ${\sf K}^{\rm Q}_n(E)$
for quenching and ${\sf K}^{\rm R}_n(E)$ for reaction given by
Eqs.(\ref{eq:K-Q}) and (\ref{eq:K-R}), respectively, for scattering of
D on initially prepared para-H$_2(v,j=0)$ and
ortho-H$_2(v,j=1)$. Here, $n=(a=1,v,j=0)$ for para-H$_2$ and
$(a=1,v,j=1)$ for ortho-H$_2$, with $a=1$ labeling the arrangement
H$_2$+D ($a=2$ stands for the arrangement HD+H).
Figure~\ref{fig:overvu-4panel} shows the overview of the results for
the six lowest vibrational levels $v$ of H$_2$.  Note that, for $v=0$,
only the reaction channel is opened for both para- and ortho-H$_2$. 
Overall, we find that ${\sf K}_v(E) $ of para- and ortho-H$_2$
share many features; for example, the position of resonances occur at
roughly the same energy, since they are due to shape resonances in the
entrance channel. However, we also notice differences, such as the
values of ${\sf K}_v(E) $ and the absence of certain resonant features
in a few cases.

\begin{figure}[t]
\centerline{
    \includegraphics[clip,width=0.75\textwidth]{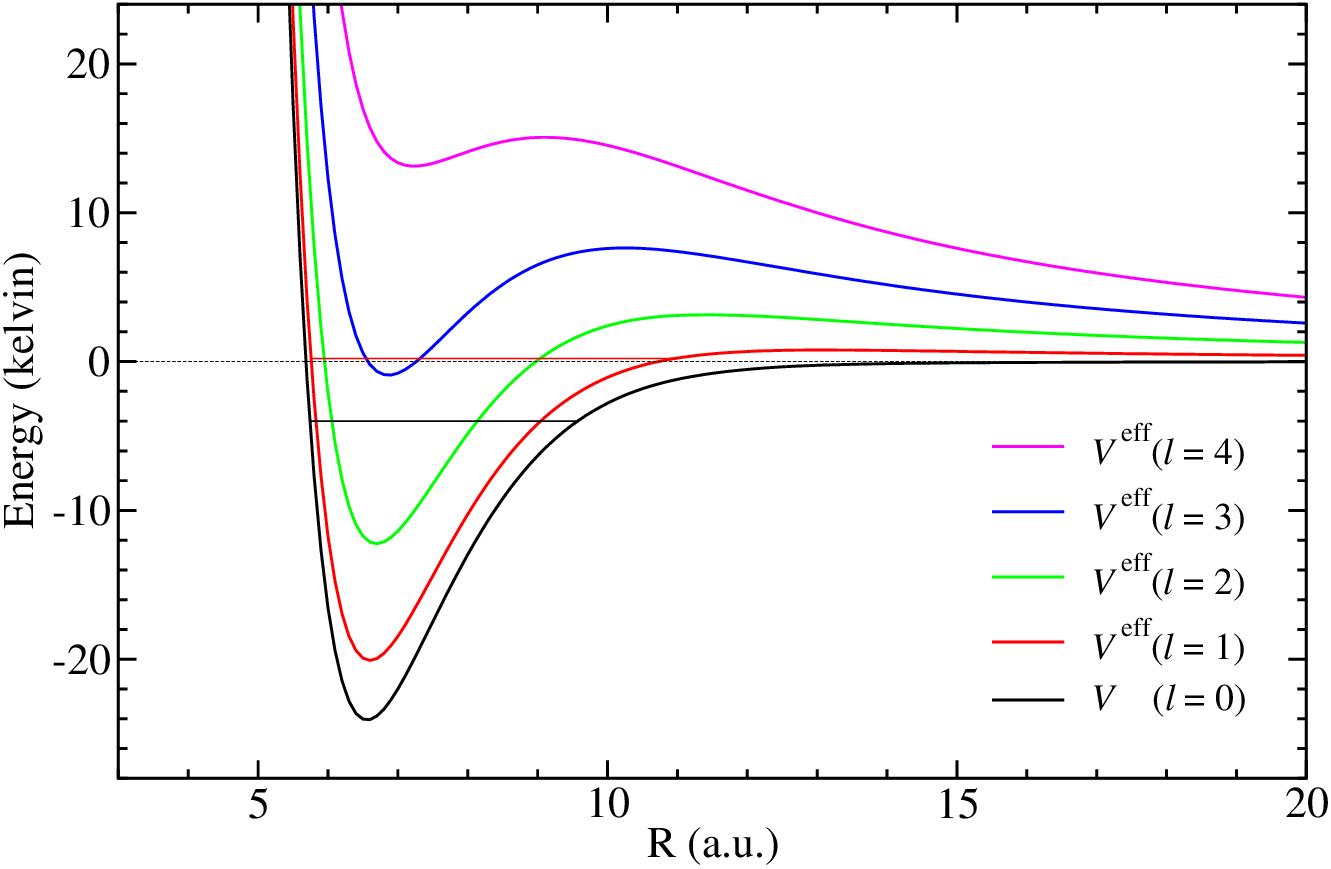}
}
\caption{Effective potential for D + para-H$_2(v=0,j=0)$
  for different partial waves $\ell$.  For $\ell=0$ and 1, the van der
  Waals complex of H$_2$+D support a bound state (indicated as a
  straight horizontal lines in both cases). As $\ell$ increases, the
  centrifugal interaction lifts the well as to prevent bound states
  and even resonances.}
\label{fig:Veff-ell}
\end{figure}

We find structures in the range of scattering energies corresponding
roughly to 100 mK and 10 K. These features are due to low partial
waves, with the more pronounced ones found at very low energies, near
15~mK, arising from $p$-wave resonances.  At higher energies (above
10~K) no structure or resonance feature can be found, because the much
higher centrifugal barrier (for higher partial waves, $\ell>3$) in the
entrance channel lifts the shallow attractive well entirely (see
Fig.~\ref{fig:Veff-ell} for D+para-H$_2$ in $v=j=0$), and thus removes
any possible bound states (i.e., no van der Waals complexes
exist for large $\ell$).  For collision energies between 10~K and
100~K, the rate constant follows a simple behavior (nearly constant).
For much higher energies (above 100~K, not shown here) an exponential
increase is typical (especially for lower vibrational initial states).
Finally, we also notice that both reaction and quenching rate
constants for ortho-H$_2$ are slightly larger than the corresponding
rates for para-H$_2$. This is to be expected, since internal energy of
ortho-H$_2$ initially in $j=1$ is slightly higher than that of
para-H$_2$ in $j=0$, thus reducing the effect of the reaction barrier,
and hence increasing the reaction and quenching processes.

In the next sections, we describe the results for both para- and
ortho-H$_2$ in more details, but we first make a general remark about
the magnitude of the rate coefficients in
Fig.~\ref{fig:overvu-4panel}.  As we discussed in our previous work
\cite{PCCP-H2+D}, tunneling through the reaction barrier for $v=0$ and
$v=1$ is responsible for the small values of
$\mathsf{K}^{\mathrm{R}}(E)$ in the energy range considered here; in
particular, for $v=1$, we have
$\mathsf{K}^{\mathrm{R}}<\mathsf{K}^{\mathrm{Q}}$.  However, for $v=2$
and higher, tunneling is no longer important; moreover, due to the
restriction on the rotational levels for para- and ortho-H$_2$, and
also because of the increased reduced mass of HD, there are many more
reaction channels $(v',j')$ for the product HD than for quenching.
Consequently, we have
$\mathsf{K}^{\mathrm{R}}>\mathsf{K}^{\mathrm{Q}}$ for $v\ge2$.

\begin{figure}[t]
\includegraphics[clip,width=0.99\textwidth]{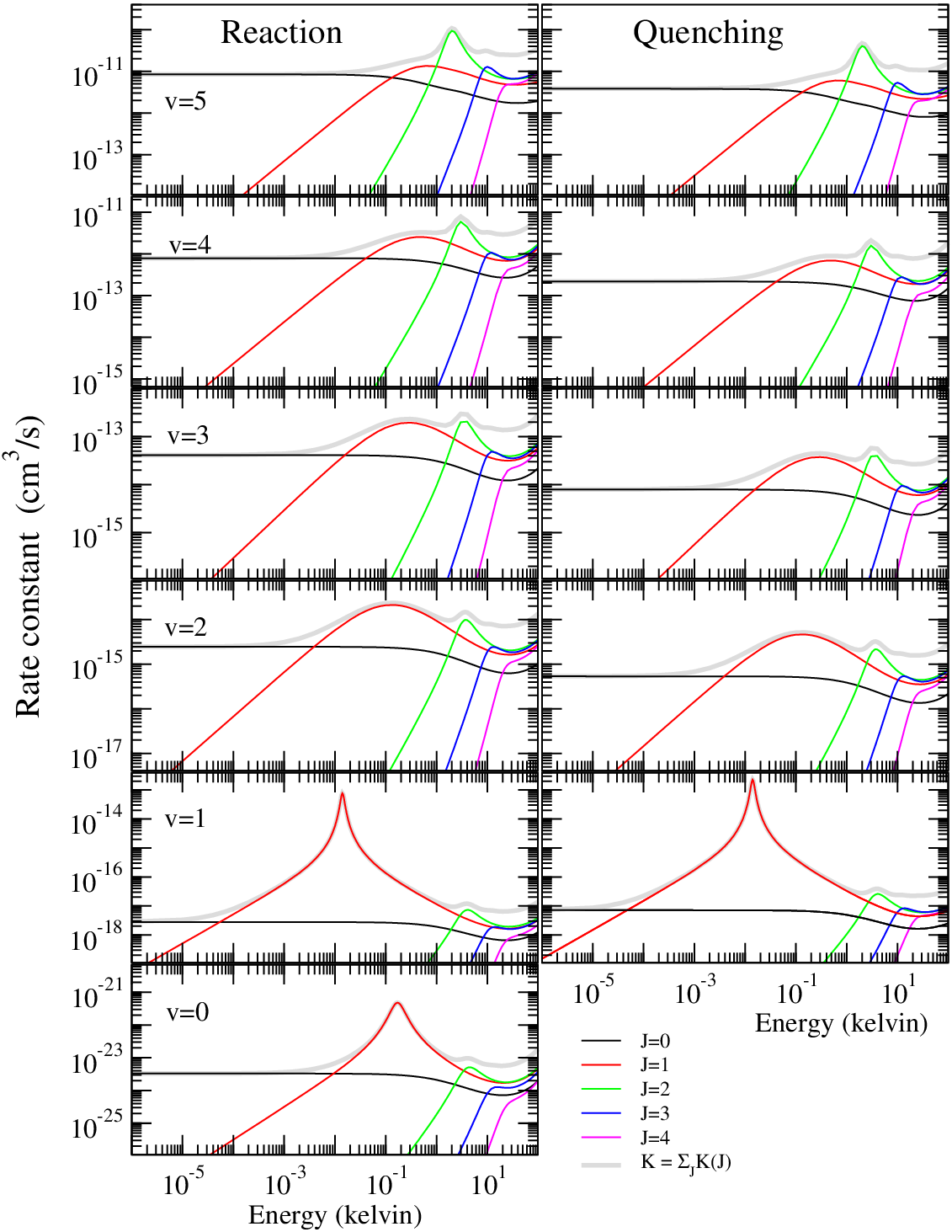}
\caption{Rate constant ${\sf K}_v(E)$ ($v\leq 5$)
  vs. collision energy corresponding to the range of 1 $\mu$K to 100 K
  for reaction (left panels) and quenching (right panels) processes
  for para-H$_2$ + D. Each panels shows the total rate constant as
  well as the contributions of individual $J=0 \dots 4$. }
\label{fig:sigma-j-2by6-para}
\end{figure}

\subsection{Para-H$_2$}
\label{sec:para}
 
In the case of para-H$_2$ with $j=0$, we have $\ell = J$ only, which
reduces the summation over $\ell$ to a single term.  The cross section
(\ref{eq:sigma}) becomes
\begin{equation}
\label{eq:sigma-para}
\sigma_{n'\leftarrow n}(E) = \frac{\pi}{k^{2}_n}
	\sum_{J=0}^{\infty} \left( 2J+1 \right)
    \;\;\sum_{\ell'= |J-j'|}^{|J+j'|}
     \left| T^{J}_{n'\ell'  nJ}(E)\right|^{2} \;.
\end{equation} 
The resulting rate constant ${\sf K}_v(E) = v_{\rm rel}\sigma$ for
both reaction and quenching processes are shown in
Fig.~\ref{fig:sigma-j-2by6-para}. For scattering energies ranging from
1 $\mu$K to roughly 100 K, only $J=0$ to 4 contribute significantly to
the cross sections and rate constants: these individual contributions
are also shown.  We now discuss each individual initial vibrational
level $v$ for both reaction and quenching.

For $v=0$, only reactions into HD($v'=0,j'\leq 2$) are possible for
this range of energies: quenching to lower states is not possible
since H$_2$ is already in its lowest state, and reactions to higher
$v'$ or even higher $j'$ would correspond to excitations and require
larger scattering energies; e.g., about 500 K to excite
H$_2(v'=0,j'=2)$ and 350 K to excite HD$(v'=0,j'=3)$. Due to the
presence of a barrier (see Fig.~\ref{fig:contour-combined}), reactions
take place through tunneling, leading to a very small rate
constant. We find a large resonant feature near 200~mK arising from
the $J=1$ contribution (or $p$-wave $\ell =1$ since $\ell =J$ for
para-H$_2$ initially in $j=0$).  As shown in Fig.~\ref{fig:Veff-ell},
the resonance occurs because the centrifugal interaction pushes the
bound state of the van der Waals complex near the scattering
threshold.  Below 10 mK, the main contribution is from $J=0$ (or
$s$-wave $\ell=0$), while $J=2$ (or $d$-wave $\ell=2$) impacts ${\sf
  K}(E)$ at roughly 3-4 K. At higher energies, the other partial waves
add up (including possible excitations), but no structure is present.

For $v>0$, both reaction and quenching processes have qualitatively
the same behaviors, with similar contributions from the different
$J$. Except for $v=1$, the reaction rate constant is larger than the
quenching rate constant, simply reflecting the larger number of exit
channels to form HD as compared to the number of channels to remain as
H$_2$, roughly 2-4 in that range of levels $v$ (see
Fig.~\ref{fig:Levels}). Level $v=1$ is slightly different because the
presence of the reaction barrier is still relevant for a sizable range
of relative orientations: as shown in Fig.~\ref{fig:contour-combined},
the reaction barrier disappears for angle larger than 112$^{\rm o}$.
The presence of the barrier for angle less than 112$^{\rm o}$ implies
a reduced reaction rate, since it requires tunneling to occur.

The $J=0$ contribution, which corresponds to $s$-wave ($\ell=J=0$)
scattering for H$_2$ in initial states $j=0$, dominates the rate
constant at very low energies, as is well known.  However, note that
for energies above 100~mK, the s-wave contribution has a minimum;
thus, higher $J$-contributions do become dominant as the scattering
energy increases. It begins with $J=1$ corresponding to $p$-wave
scattering ($\ell=J=1$), followed by $J=2$ ($d$-wave scattering with
$\ell=J=2$), and so on. The exact energy at which these higher $J$
contributions become important depends on the energy surface and the
position of levels.

For both $v=0$ (reaction only) and $v=1$, the main feature is a
scattering resonance due to $J=1$ ($p$-wave scattering with
$\ell=J=1$), arising from a quasi-bound state in the van der Waals
H$_2\cdots$D complex. From $v=0$ to $v=1$, the quasi-bound level gets
closer to the threshold, leading to a shift of the resonance to a
lower scattering energy (from 200~mK for $v=0$ to 15~mK for $v=1$).
This quasi-bound level becomes bound for $v=2$, removing the resonant
feature from the rate constant, and becomes more deeply bound as $v$
increases, moving the $J=1$ contribution to higher scattering
energies. The $J=2$ contribution becomes important at higher energies
in the range of a few K, and shifts slightly to lower scattering
energies as $v$ grows; it starts to be important for $v=2$, and
becomes the key feature at higher $v$. Similarly, $J=3$ ($f$-wave
scattering with $\ell=J=3$) starts contributing at roughly 10 K, with
the exact position of its maximum contribution shifting slightly with
$v$; it leads to a perceptible feature only for the highest $v$, its
contribution being otherwise a simple addition to all higher $J$
giving the total rate constant. This is illustrated by $J=4$ ($g$-wave
scattering with $\ell=J=4$) which simply adds up to the total rate
constant without leading to features.

\begin{figure}[t]
\includegraphics[clip,width=0.99\textwidth]{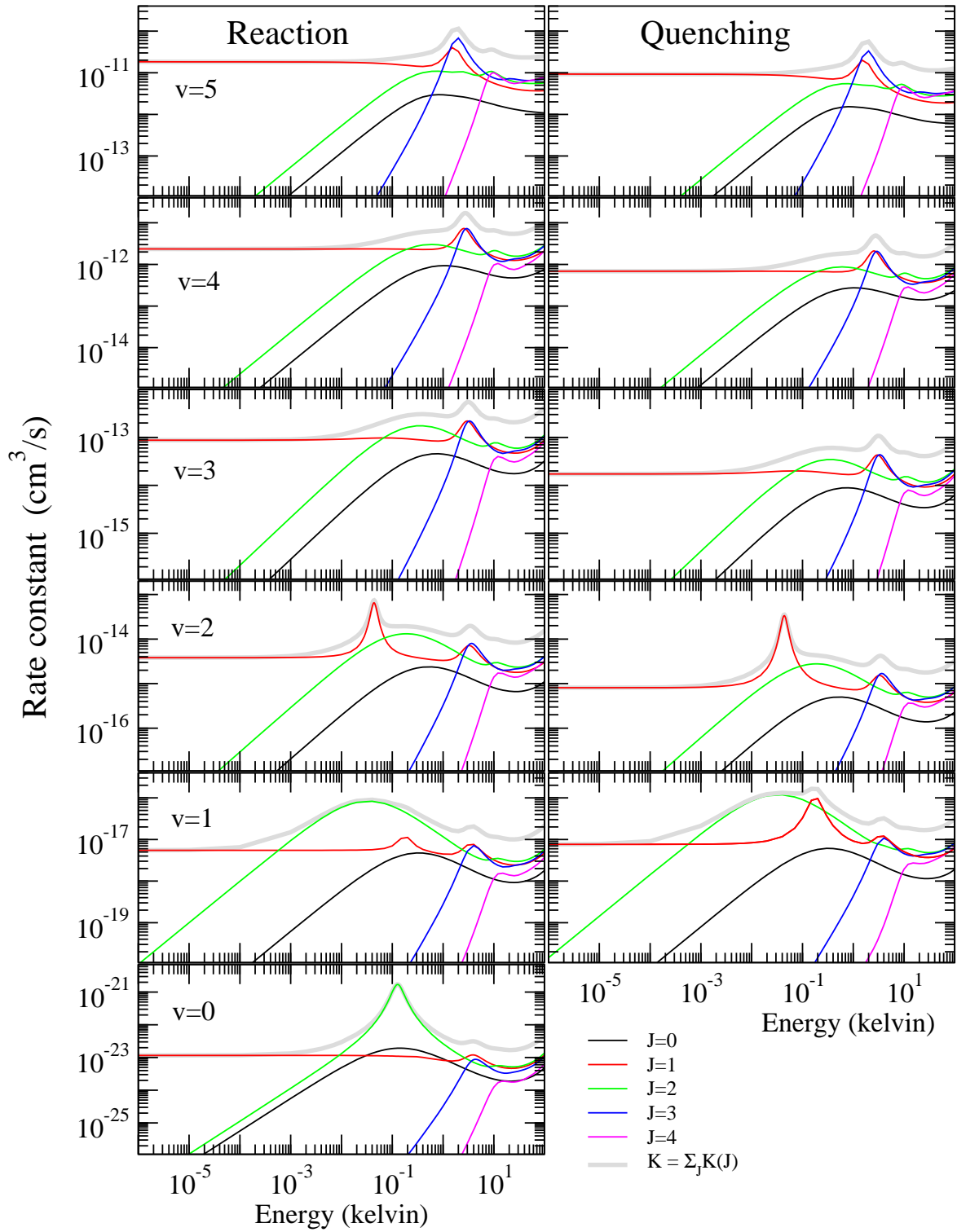}
\caption{Rate constant ${\sf K}_v(E)$ ($v\leq 5$)
  vs. collision energy corresponding to the range of 1 $\mu$K to 100 K
  for reaction (left panels) and quenching (right panels) processes
  for ortho-H$_2$ + D. Each panels shows the total rate constant as
  well as the contributions of individual $J=0 \dots 4$. }
\label{fig:sigma-j-2by6-ortho}
\end{figure}

\subsection{Ortho-H$_2$}
\label{sec:ortho}

In the case of ortho-H$_2$ with $j=1$, since $\ell = |J-j|, \cdots,
J+j$, three partial waves contribute to a given $J$, except for
$J=0$. In fact, for $J=0$, we have only the $\ell = j =1$ $p$-wave
scattering contribution, while $J=1$ includes the $\ell =0 ,1 $ and 2
($s$, $p$, and $d$-waves) contributions, $J=2$ the $\ell = 1, 2,$ and
3 ($p$, $d$ and $f$-waves) contributions, and so on.  The cross
section (\ref{eq:sigma}) becomes
\begin{equation}
\label{eq:sigma-ortho}
\sigma_{n'\leftarrow n}(E) = \frac{\pi}{k^{2}_n}
	\sum_{J=0}^{\infty} \left(\frac{2J+1}{3}\right)
	\sum_{\ell = |J-1|}^{|J+1|}
    \;\;\sum_{\ell'= |J-j'|}^{|J+j'|}
     \left| T^{J}_{n'\ell'  n\ell}(E)\right|^{2} \;.
\end{equation}
The resulting rate constant ${\sf K}_v(E) = v_{\rm rel}\sigma$ for
both reaction and quenching processes are shown in
Fig.~\ref{fig:sigma-j-2by6-ortho}. Many of the features and behaviors
described for para-H$_2$ apply for ortho-H$_2$ as well; only $J=0$ to
4 (also individually shown) contribute significantly to the cross
sections and rate constants for scattering energies ranging from 1
$\mu$K to roughly 100 K (recall that the hyperfine interactions
omitted here will become relevant below roughly 15~mK). For the same
reason as for para-H$_2$, $v=0$ leads to reaction only, since
quenching to lower states of H$_2$ is not possible (para-ortho mixing
due to nuclear spin-rotation coupling is extremely weak and is
neglected here), and the value is very small because the reaction must
occur via tunneling through a barrier.  Similarly, except for $v=1$
for which the reaction barrier still influences the rate, the reaction
rate constant is larger than the quenching constant for $v>1$.

However, the exact features are different from those of
para-H$_2$. This can be understood in terms of the different couplings
between the various states. We now describe in more detail the results
for various initial $v$. Except for $J=0$, all other $J$` include
contributions from three initial partial waves $\ell$. However, even
if a given partial wave $\ell$ participates in various
$J$-contributions, their effect is slightly different since each total
angular moment $J$ leads to different coupling strengths. For example,
the $p$-wave $\ell=1$ initial partial wave occurs in the contributions
of $J=0$, 1, and 2. For $v=0$ (reaction only), the $p$-wave resonance
appears only in the $J=2$ contribution, while the $p$-wave scattering
corresponds to a bound state in $J=0$, and has almost no influence
over the $s$-wave dominated $J=1$ contribution.  We note that the
higher $J$ are usually favored because of the $2J+1$ factor
appearing in the expression for the cross section
(\ref{eq:sigma-ortho}). Still for $v=0$, the $s$-wave scattering is
present only in the $J=1$ contribution, while the higher $J=3$ and 4
contain $\ell=2,3$, and 4, and 3, 4 and 5, respectively.  As in
para-H$_2$, the effect of partial waves higher than $d$-wave ($\ell
=2$) is hidden in the total rate; they do not produce noticeable
structures.

For all other $v$, the $J=2$ does not contain a $p$-wave resonance;
traces of it appears only in the $J=1$ contribution of $v=1$ and
$2$. In more detail, for both $v=1$ and $2$, $J=0$ includes only
$p$-wave scattering without resonance. Also, $J=1$ contains the $s$,
$p$, and $d$-wave scattering; the position of the maximum for $p$-wave
is shifted relative to the other $J$, illustrating the sensitivity
of the position of the bound level in the van der Waals complex to the
couplings. This is less so for the other partial waves since the bound
states are not close to the threshold. The $J=2$ contribution includes
results from $p$, $d$, and $f$-scattering, with the $\ell=1$ and 3
components being dominant, while the $l=2$ is obscured by them. This
occurs for all $v$. As for $v=0$, the higher contributions $J=3$ and
4 are dominated by $\ell =2$ and $\ell =3$, respectively, with signals
from $\ell =3$ and higher being hidden in the total rate constant. For
$v>2$, the $p$-wave feature of $J=1$ is absent, preventing a
double-peak structure in the total rates.

\subsection{Elastic cross sections}
\label{sec:elastic}

Here, we describe very briefly the results for the elastic cross
section, which reads
\begin{equation}
\label{eq:sigma-elastic}
\fl  \sigma^{\rm el.}_{n}(E) \equiv \sigma^{\rm el.}_{n\leftarrow n}(E) 
      = \frac{\pi}{k^{2}_n}
	\sum_{J=0}^{\infty} \left(\frac{2J+1}{2j+1}\right)
	\sum_{\ell' = |J-j|}^{J+j} \;\; \sum_{\ell = |J-j|}^{J+j}
     \left| \delta_{\ell'\ell} - S^{J}_{n\ell'\; n\ell}(E)\right|^{2}.
\end{equation}
For para-H$_2$ initially in $j=0$ (with $\ell =J$) we have 
\begin{equation}
\label{eq:sigma-elastic-para}
    \sigma^{\rm el.-para}_{n}(E) = \frac{\pi}{k^{2}_n}
	\sum_{J=0}^{\infty} \left( 2J+1\right)
     \left| 1 - S^{J}_{nJ\; nJ}(E)\right|^{2},
\end{equation}
and for ortho-H$_2$ initially in $j=1$ ( with $\ell = |J-1|,\cdots, J+1$)
\begin{equation}
\label{eq:sigma-elastic-ortho}
    \sigma^{\rm el.-ortho}_{n}(E) = \frac{\pi}{k^{2}_n}
	\sum_{J=0}^{\infty} \left(\frac{2J+1}{3}\right)
	\sum_{\ell' = |J-1|}^{J+1}  \;\;  \sum_{\ell = |J-1|}^{J+j}
     \left| \delta_{\ell'\ell} - S^{J}_{n\ell' \; n\ell}(E)\right|^{2}.
\end{equation}

\begin{figure}[t]
\includegraphics[clip,width=0.99\textwidth]{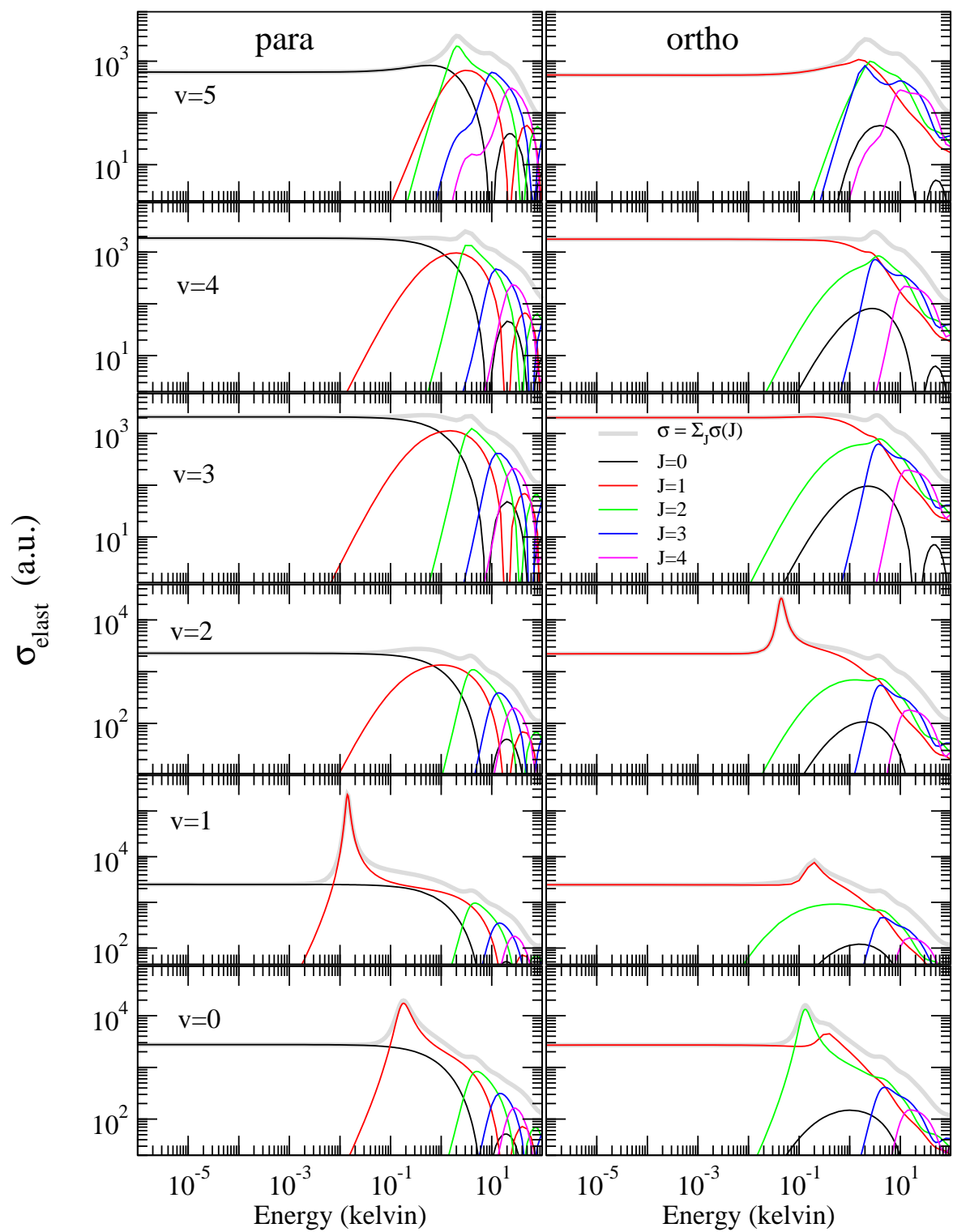}
\caption{Elastic cross sections $\sigma^{\rm el.}_v(E)$
  ($v\leq 5$) vs. collision energy corresponding to the range of 1
  $\mu$K to 100 K for D + para-H$_2$ (left panels) and ortho-H$_2$
  (right panels).  Each panels shows the total rate constant as well
  as the contributions of individual $J=0 \dots 4$. }
\label{fig:sigma-j-elast}
\end{figure}

\begin{figure}[t]
\centerline{
    \includegraphics[clip,width=0.75\textwidth]{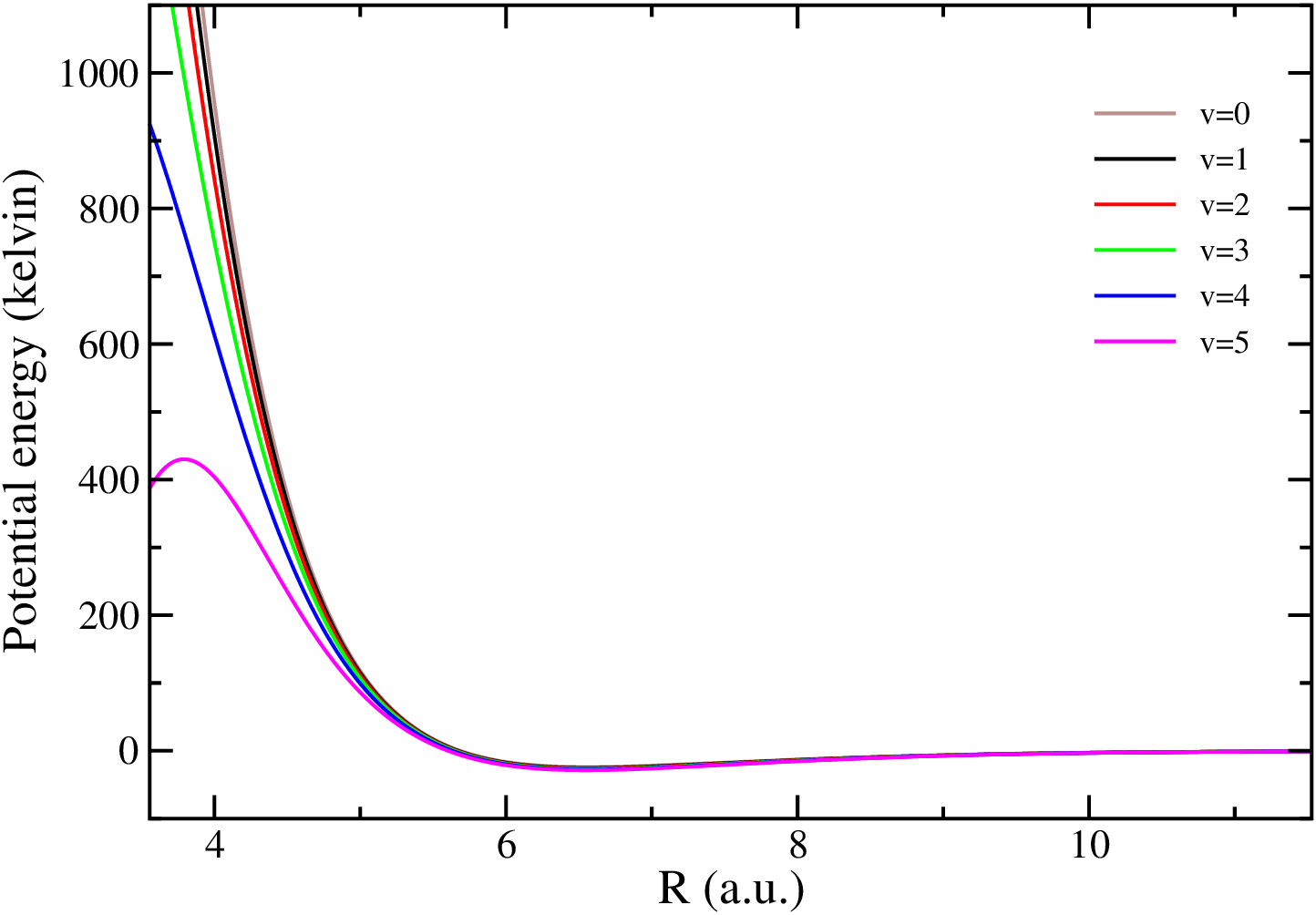}
}
\caption{Effective potential for $\ell=0$ for
  D + H$_2(v,j=0)$. The asymptotes of the curves for the different
  levels $v$ are all set to zero. The curves are very similar at large
  separation, and start to become different at short range for higher
  $v$. The difference becomes sizable at energies corresponding to
  roughly 200 K.}
\label{fig:vpot}
\end{figure}

The corresponding results are shown in Fig.~\ref{fig:sigma-j-elast}
for $v\leq 5$.  As before, there is some structure in the range of
roughly 10 mK to about 10 K.  The cross sections for both para- and
ortho-H$_2$+D have the same order of magnitude for all $v$, except
in cases for which the $p$-wave scattering leads to a sharp
feature. This occurs for $v=0$ and 1 of both para- and ortho-H$_2$, and
for $v=2$ of ortho-H$_2$. Overall, the elastic cross sections have
basically the same value around 2000 $a_0^2$, with very small
variation about that value. This is expected, since the entrance
channels are very similar for all those initial states (see
Fig.~\ref{fig:vpot}), as opposed to the reaction and quenching
processes where the position of the exit channels and the coupling
strengths dictate the large range of constant rate values.

\subsection{State-to-state results}

In this section, we give an overview of the richness of the
state-to-state processes.  We first discuss the branching ratios of
product formation from each initial vibrational state $v$ of H$_2$
into individual vibrational levels $v'$ (summed over $j'$) for
quenching and reaction separately, together with the branching ratio
into all product channels (i.e., quenching and reaction
combined). This is followed by an example of a typical case for
rotationally-resolved rate constants. In all cases, we contrast the
results of para- and ortho-H$_2$.

\subsubsection{Products branching ratios.}

\begin{figure}[t]
\includegraphics[clip,width= 0.99\textwidth]{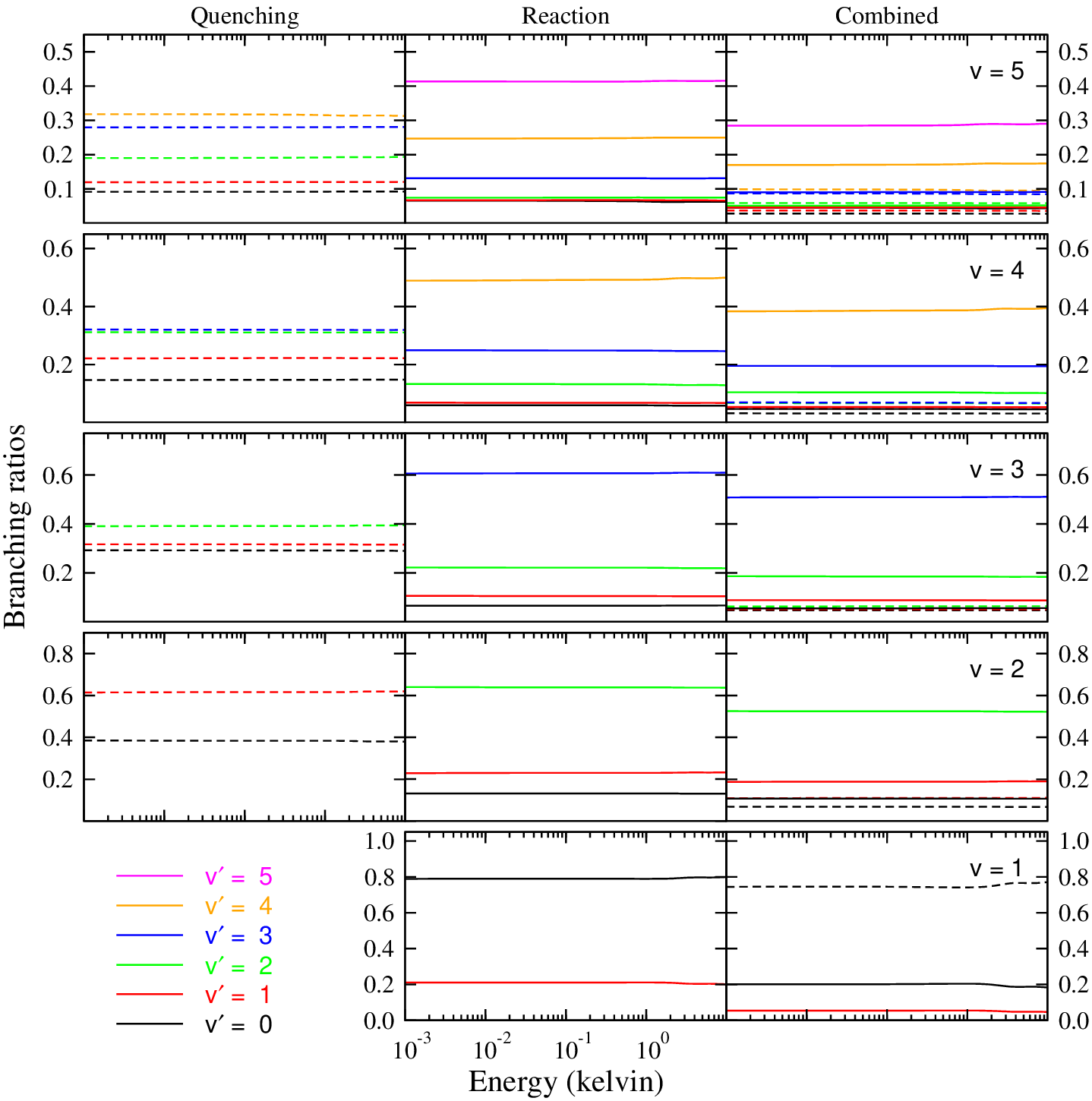}
\caption{
   Branching ratios of product formation for D + para-H$_2(v,j=0)$ into
   individual vibrational levels $v'$ (summed over $j'$) for quenching
   of H$_2$ only (left panels: dashed lines) and reaction forming HD
   only (middle panels: solid lines). The right panels show the
   branching ratios when both quenching and reaction are combined
   (quenching: dashed lines, reaction: solid lines). The initial $v=0$
   panels and the panel for $v=1$ of quenching alone are omitted (see
   text).
}
\label{fig:branch-para}
\end{figure}

We first give the branching ratios for para-H$_2$ initially in the
$(v,j=0)$ state.  They are depicted in Fig.~\ref{fig:branch-para}; the
left panels show ratios for quenching alone (dashed lines), the middle
panels for reaction alone (solid lines), and the right panels for the
combined inelastic processes (quenching: dashed lines, reaction: solid
lines).  We omit the initial $v=0$ panels since there is no quenching
possible and all reactions end up in $v'=0$ of HD, and the panel for
$v=1$ of quenching alone, since all quenching goes into $v'=0$ of
H$_2$ in this particular case.

We find that the branching ratios are basically insensitive to the
scattering energy within the 1 mK to 10 K range, even though there is
structure in the corresponding rate constant shown in
Fig.~\ref{fig:sigma-j-2by6-para}. This simply signifies that all the
flux into the various product channels are changing in unison with the
appearance of scattering features; these are due to shape resonances
and higher partial waves in the entrance channel that raise all
scattering amplitudes in the same proportion. In detail, starting from
$v=5$, the largest fraction (32\%) of quenching is in the uppermost
level $v'=4$, followed by 28\% in $v'=3$, 19\% in $v'=2$, 12\% in
$v'=1$, and 9\% in $v'=0$. For reaction alone, $v'=5$ dominates the
reaction products formed with 41\%, followed by 25\% in $v'=4$, 14\%
in $v'=3$, while each of $v'=2$, 1, and 0 have between 6\% and 8\%.
When all product channels are combined, it becomes clear that the most
probable products stem from the reaction forming HD in $v'=5$ (29\%),
and 4 (17\%), with the rest of reaction and quenching being comparable
at 10\% or less.  As discussed in Section~\ref{sec:statistical} below,
this is to be expected, since the number of possible exit channels is
larger in the reaction arrangement than in the quenching arrangement
\cite{PCCP-H2+D}.

The same general behavior takes place for $v>1$: for $v=4$, both
$v'=3$ and 2 have the same quenching ratios ($\sim$32\%), while $v'=4$
leads the reaction ratios with 50\%, and the most probable products
are HD in $v'=4$ (39\%) and 3 (19\%), the remaining being distributed
over the other channels. For $v'=3$, quenching is largest into $v'=2$
(39\%), reaction into $v'=3$ (60\%) and 2 (22\%), and the most
probable products are HD in $v'=3$ (51\%) and 2 (18\%). Similarly for
$v=2$, 62\% of the quenching ends up in $v'=1$ and 38\% in 0, while
64\% of reaction goes into $v'=2$, 23\% in 1, and 13\% in 0: most of
the products are HD in $v'=2$ (52\%) and 1 (19\%), with the rest
distributed over the remaining channels.

Now, the behavior for $v=1$ reflects the considerable effect of the
barrier on the scattering: 80\% of the reaction into HD produces the
lowest level $v'=0$ as compared to 20\% in the highest open level
$v'=1$.  The reaction fractions are much smaller than the quenching
fraction; specifically, 74\% for quenching into H$_2(v'=0)$, 20\% and
6\% into $v'=0$ and 1 of HD, respectively. Finally, for $v=0$ (not
shown in the figure), only reaction into $v'=0$ of HD is possible.

In general, we find that the products with the largest (although not
dominant) branching ratios are into highest $v'$ allowed: $v'=v$ for
reaction, and $v'=v-1$ for quenching. Also, reaction products are more
likely than quenching, which is expected since there are more possible
exit channels available in the HD reaction arrangement as compared to
the H$_2$ quenching arrangement. These conclusions apply to all
initial states $v$, except when tunneling through the barrier is
important, as discussed in Section~\ref{sec:statistical}.

\begin{figure}[t]
\includegraphics[clip,width= 0.99\textwidth]{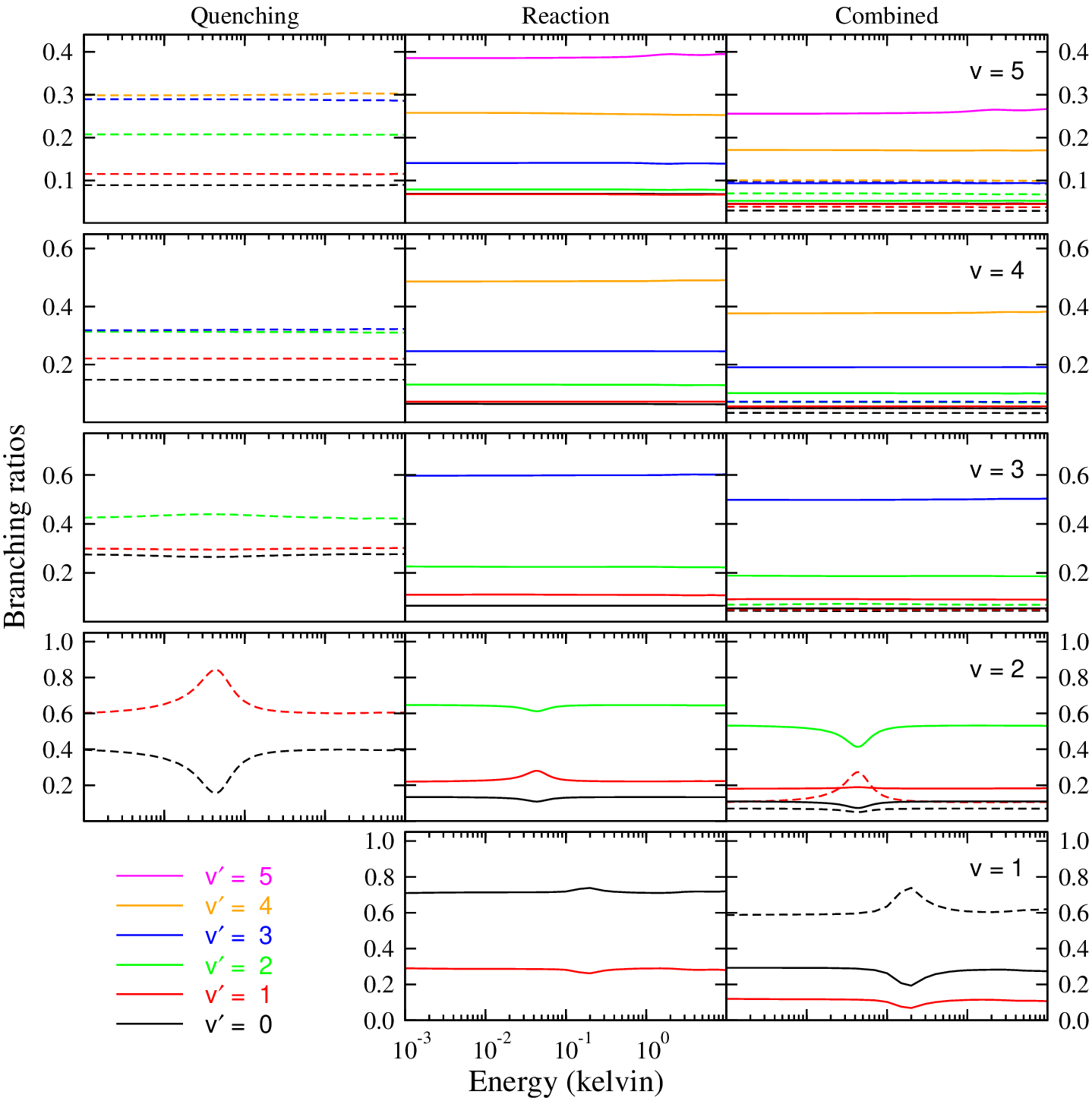}
\caption{Same as Fig.~\ref{fig:branch-para} for D + ortho-H$_2(v,j=1)$.}
\label{fig:branch-ortho}
\end{figure}

The same overall conclusions apply to the case of ortho-H$_2$
initially in $(v,j=1)$ shown Fig.~\ref{fig:branch-ortho}: in general,
the branching ratios are rather insensitive to the scattering energy
in the range of 1~mK to 10~K, the largest branching ratios are into
products with the highest $v'$ allowed (with reaction to form HD more
probable than quenching of H$_2$), and tunneling through the barrier
affects the lowest initial levels $v=1$ and 0. However, we note that
the branching ratios are changing for both $v=2$ and 1. In $v=2$, the
quenching into $v'=1$ increases from 60\% to 85\% near 50 mK, while
quenching into 0 decreases accordingly from 40\% to 15\%.  For the
reaction products, the branching ratios into $v'=2$ and 0 decrease
slightly at the same energy, while it increases into $v'=1$. The net
effect in the combined branching ratios shows a sizable increase of
the quenching into $v'=1$ (reaching 25\%) while reaction into $v'=2$
decreases to 40\%. For the initial $v=1$ state of H$_2$, the net
result is an increase of the branching ratio into H$_2(v'=0)$ from
60\% to 75\% near 200 mK, while the reaction into HD in $v'=0$ and 1
are both decreasing accordingly.  This is an example showing that
para- and ortho-H$_2$ have different behaviors. To better understand
the origin of those variations, we investigate rotationally-resolved
rate constants for a specific case, namely $v=2$, in the next section.

\subsubsection{Rotationally-resolved rate constants for H$_2$ in $v=2$.}

Because of the large number of exit channels for a given entrance
channel, we restrict this analysis to a typical entrance channel
showing structure, while still having a manageable amount of exit
channels. We selected to show the state-to-state inelastic (reaction
and quenching) rate constant for para- and ortho-H$_2$ initially in
$v=2$. Other entrance channels would exhibit the same type of
behavior, though the exact details will vary. However, as will become
obvious below, the rapidly growing number of channels make it
difficult to show those results.

\begin{figure}[t]
\includegraphics[clip,width= 0.99\textwidth]{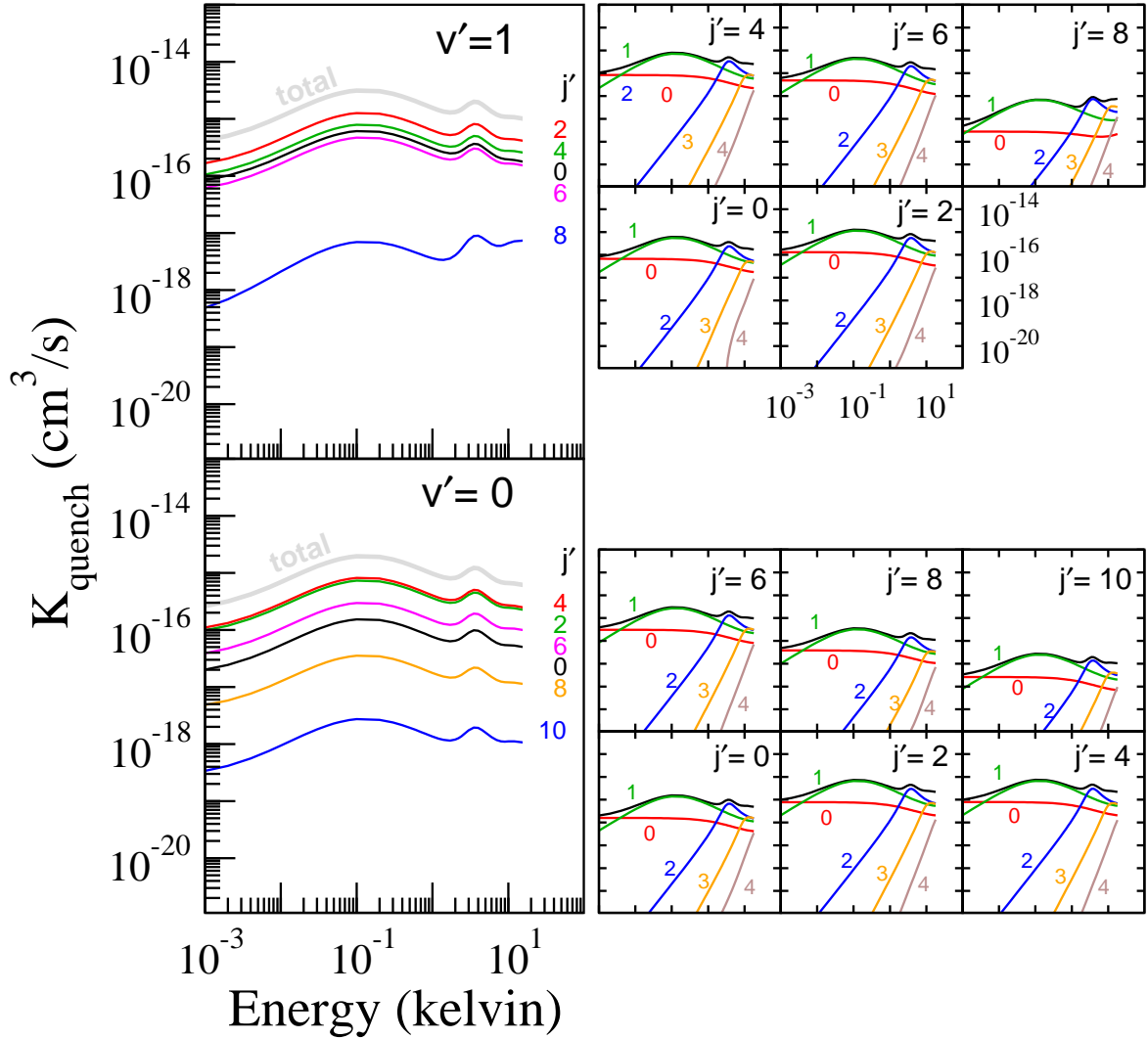}
\caption{Rate constant ${\sf K}(E)$ for para-H$_2$ ($v=2$) for
  quenching into specific exit channels $v'$ (left panels). Each shows
  the contributions of individual rotational states $j'$. For each
  value of $j'$, the contributions of each $J=0 \dots 4$ are shown
  (right panels).}
\label{fig:state-para-quench}
\end{figure}

\begin{figure}[t]
\includegraphics[clip,width= 0.99\textwidth]{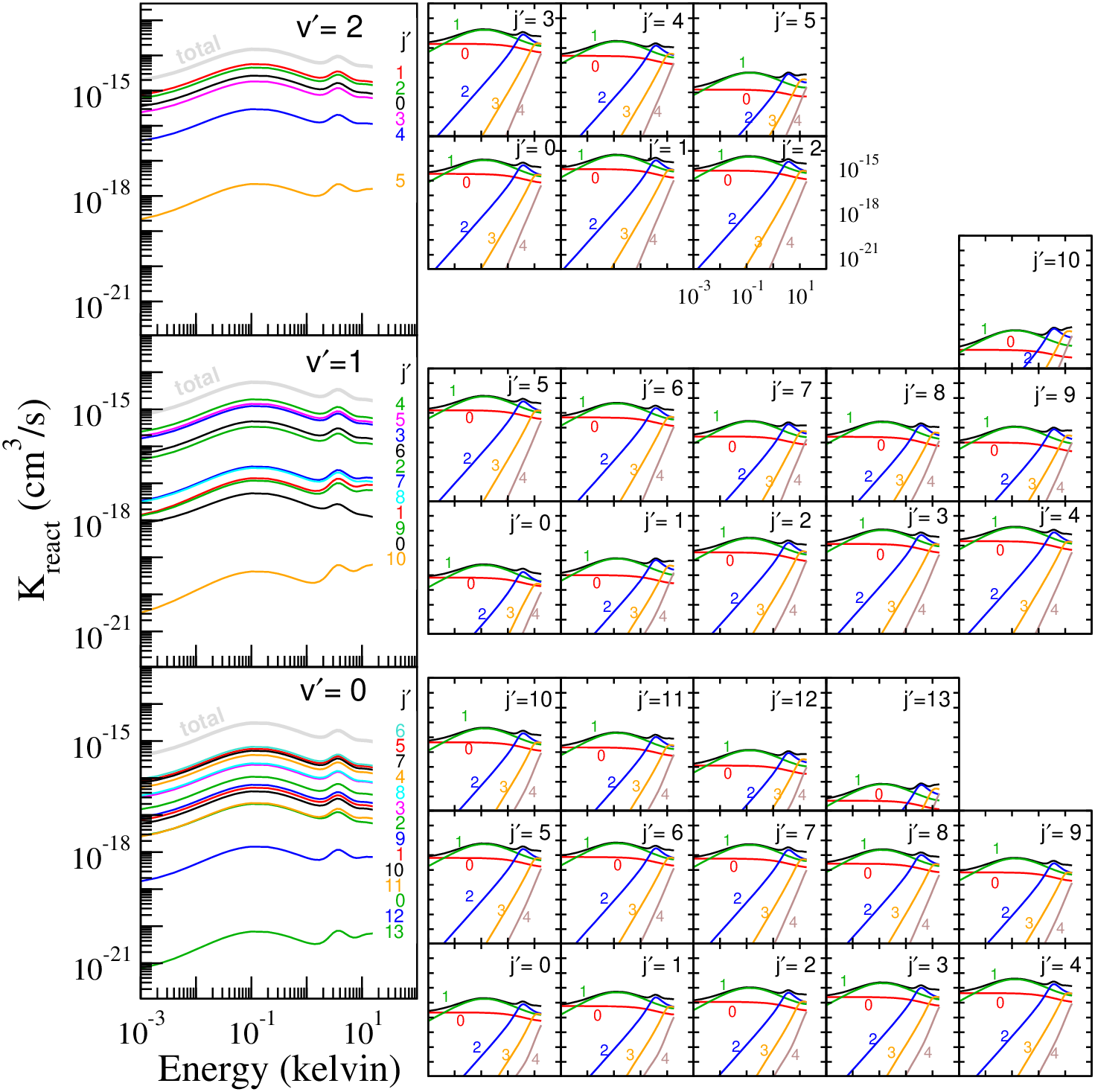}
\caption{Same as Fig.~\ref{fig:state-para-quench} for
              reaction. The quantum numbers $v'$ and $j'$ label the
              rovibrational channels of the product HD.}
\label{fig:state-para-react}
\end{figure}

We begin with para-H$_2$ initially in $(v=2,j=0)$, and show the
state-to-state rate constant for quenching in
Fig.~\ref{fig:state-para-quench}, and for reaction in
Fig.~\ref{fig:state-para-react}.  For quenching, we show the results
in $v'=1$ (top left panel) and $v'=0$ (bottom left panel): in each of
exits levels, we show the individual contributions of rotational
states $j'$ to the total rate constant (with the quantum numbers
labels following the same ordering as the partial rate constants). To
the right of those two panels, we show the contribution of individual
$J$ total angular momenta for each channel $j'$. In all cases, all
axes have the same range. For $v'=1$, the most important contribution
comes from $j'=2$, followed by 4, 0, 6, and 8: each have a very
similar energy dependence.  The individual contribution of each $J$
are shown in the right panel: they all have the same energy
dependence, with the main contribution arising from $J=1$ and 2 (at
higher energies).  The results for $v'=0$ are very similar, although
the ordering of the $j'$ contributions changes a bit and an additional
$j'=10$ appears.

For reaction (see Fig.~\ref{fig:state-para-react}), very similar
behaviors are found, although the number of exit channels $v'$ now
include 2 as well, and both odd and even $j'$ are allowed. The exact
ordering of contributions differs from the quenching case, but the
overall system exhibit the same dependence.

\begin{figure}[t]
\includegraphics[clip,width= 0.99\textwidth]{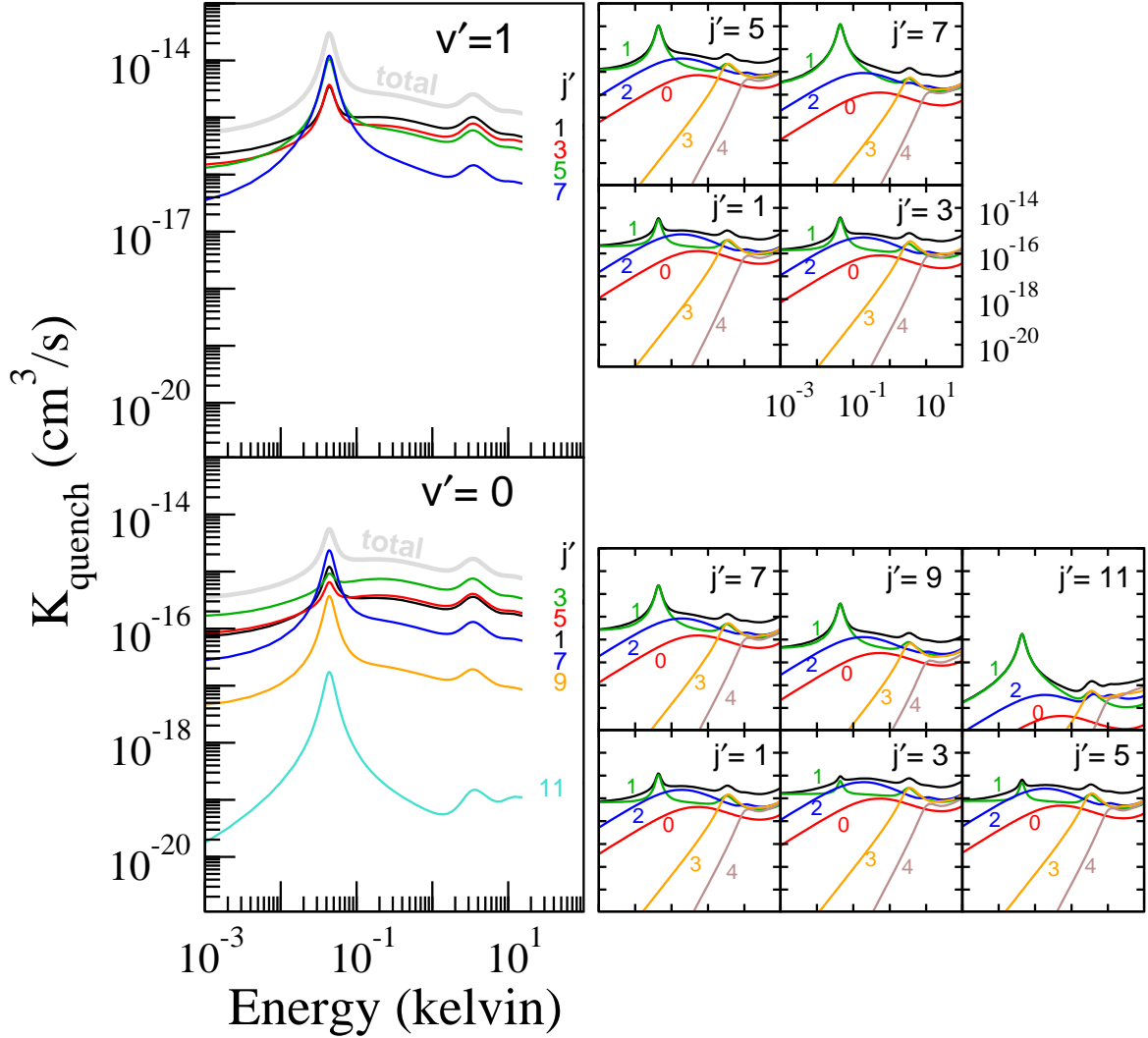}
\caption{Rate constant ${\sf K}(E)$ for ortho-H$_2$ ($v=2$) for
  quenching into specific exit channels $v'$ (left panels). Each shows
  the contributions of individual rotational states $j'$. For each
  value of $j'$, the contributions of each $J=0 \dots 4$ are shown
  (right panels). }
\label{fig:state-ortho-quench}
\end{figure}

\begin{figure}[t]
\includegraphics[clip,width= 0.99\textwidth]{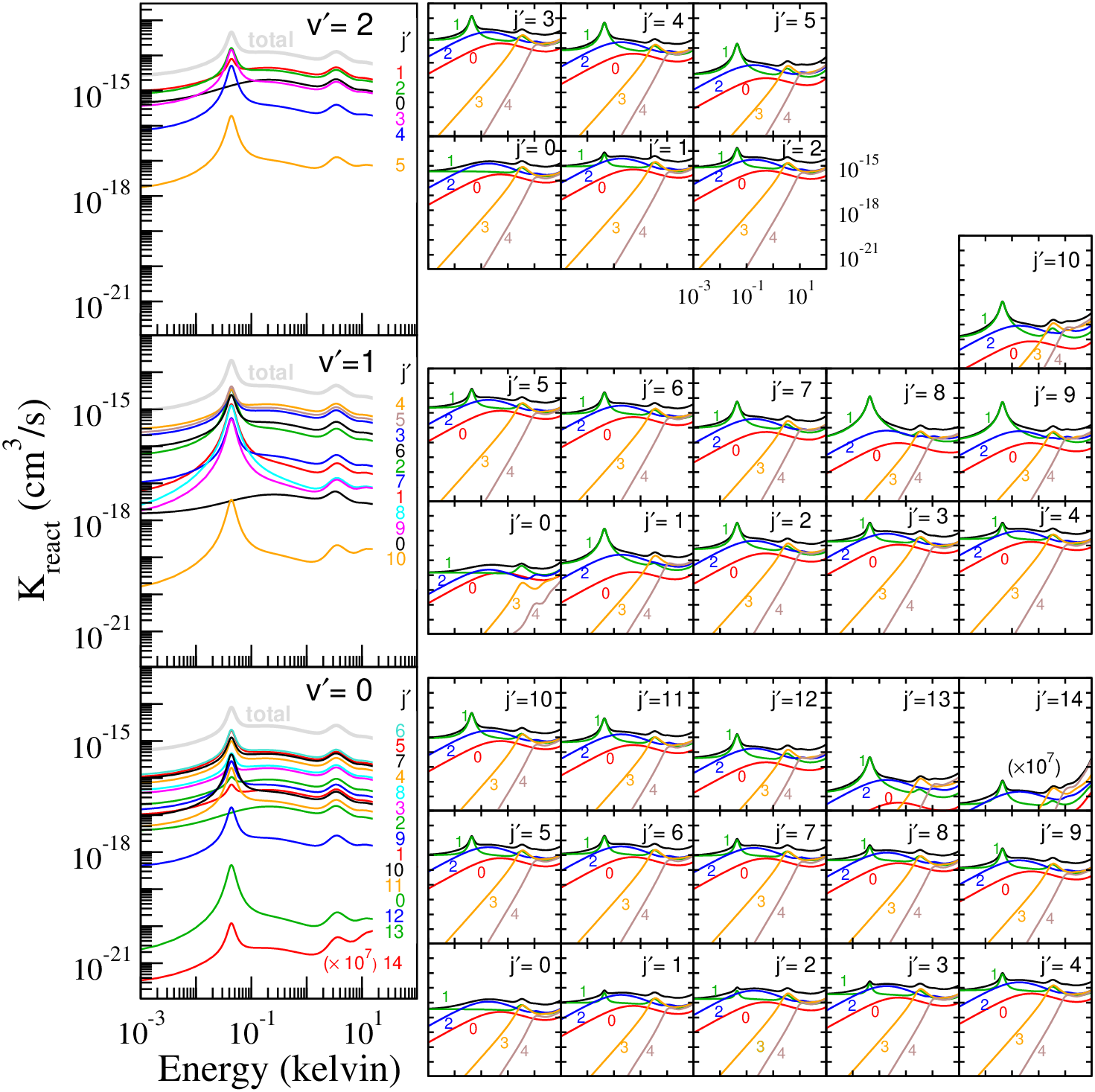}
\caption{Same as Fig.~\ref{fig:state-ortho-quench} for
              reaction.}
\label{fig:state-ortho-react}
\end{figure}

The same information for the case of ortho-H$_2$ initially in
$(v=2,j=1)$ is shown in Fig.~\ref{fig:state-ortho-quench} for
quenching and Fig.~\ref{fig:state-ortho-react} for reaction. For
quenching alone, the left panels in Fig.~\ref{fig:state-ortho-quench}
depict the rate constant for $v'=1$ and 0, each for the various odd
$j'$ contributions to the total value. Although these contributions
show similar dependence on the scattering energy for both $v'$, one
can notice that specific $j'$ have larger increases than others near
the resonant feature at 50 mK. For example, in $v'=1$, $j'=7$ becomes
dominant at the resonance, even though it is the
smallest contribution away from it. The same is true to a lesser
extent for the other $j'$. This is more apparent in $v'=0$, where not
only $j'=7$ (which becomes dominant) but also 9 and 11 have sharper
increases at the resonant structure. The origin of those variations is
due to the different couplings between the various states, as
discussed in Section~\ref{sec:ortho}. Basically, this feature arises
from a resonance in the $p$-wave ($\ell=1$) initial partial wave,
which occurs in the contributions of $J=0$, 1, and 2, and their effect
is slightly different since each total angular moment $J$ leads to
different coupling strengths.

For reaction alone, shown in Fig.~\ref{fig:state-ortho-react}, we have
a very similar story, though there are many more possible product
states (including $v'=2$, and even and odd $j'$). Again, the structure
is due to the $p$-wave in the entrance channel, and the contributions
of the various $j'$ increase different near the resonant feature at 50
mK. For example, for $v'=2$, $j'=3$ and 4 have sharper increases,
while the same is true for $j'= 2,6$ and 8 of $v'=1$. These ``unequal"
changes in the contributions of the total rate constant in each
product $v'$ explains the difference between the state-to-state
scattering of para- and ortho-H$_2$.

Rotational product distributions for a given scattering energy $E$,
i.e., the probabilities $p^{n}_{a',v'}(j',E)$ to form a product in a
rotational state $j'$ (in vibrational state $v'$) in arrangement $a'$
can be obtained by simply dividing the rate constant for this specific
channel by the total rate constant for the corresponding product
channel $v'$, namely $p^{n}_{a',v'}(j',E) = {\sf
  K}_{a',v',j'\leftarrow n} (E)/\sum_{j'}{\sf K}_{a',v',j'\leftarrow
  n}(E)$, where $n=(a,v,j)$ stands for the entrance channel $(v,j)$ in
the reactant arrangement $a$.  By summing over all open exit channel
within one arrangement $a'$, one can find the rovibrational
distribution for quenching ($Q$: $a'=a$) or reaction ($R$: $a'\neq
a$), or simply $p^{n}_{Q(R)}(v',j',E) = {\sf K}_{Q(R),v',j'\leftarrow
  n} (E)/{\sf K}_n^{Q(R)}(E)$.  Finally, the distribution over all
  possible channels can be obtained by dividing ${\sf
    K}_{Q(R),v',j'\leftarrow n} (E)$ by total rate constant ${\sf
    K}_n(E)={\sf K}_n^{Q}(E)+{\sf K}_n^{R}(E)$.

\subsection{Statistical model}
\label{sec:statistical}

Scattering problems with rearrangement are computationally demanding
when a full quantum mechanical approach is used; thus, approximate
models have been developed, often based on statistical formalism and
semi-classical treatments \cite{miller,dulieu}.  Here, we compare the
results of our full quantum computation with a statistical treatment
based on the phase-space theory (PST) \cite{miller}.

Because the entrance channel plays a special role at low scattering
energies, it is advantageous to interchange the order of the $J$ and
$\ell$ summations in Eq.(\ref{eq:sigma}), and rewrite the cross
section as
\begin{equation}
\label{eq:sigma-reversed}
\sigma_{n'\leftarrow n}(E) = \frac{\pi}{k^{2}_n}
    \sum_{\ell=0}^\infty \sum_{J=|\ell-j|}^{\ell +j}
	\left(\frac{2J+1}{2j+1}\right)
    \;\;\sum_{\ell'= |J-j'|}^{|J+j'|}
     \left| T^{J}_{n'\ell'  n\ell}(E)\right|^{2},
\end{equation}
Following Miller \cite{miller}, we can write the $T$-matrix element
$T^{J}_{n'\ell'  n\ell}$ for an inelastic process (i.e., $n\neq n'$) as
\begin{equation}
\label{eq:T-PST}
 |T^{J}_{n'\ell'  n\ell}(E)|^2 
= \frac{P^J_{n,\ell}(E) P^J_{n',\ell'}(E)}{\sum_{n'',\ell''}P^J_{n'',\ell''}(E)},
\end{equation}
where $P^J_{n,\ell}(E)$ is the energy-dependent capture probability in
the channel $n$ for a given partial wave $\ell$ of a particular total
angular momentum $J$, with similar definitions for $P^J_{n',\ell'}(E)$
and $P^J_{n'',\ell''}(E)$. Labeling the initial arrangement H$_2$+D by
$a$, the rate constant ${\sf K}_{n'\leftarrow n}(E) = v_{\rm rel.}
\sigma_{n'\leftarrow n}(E)$ with $v_{\rm rel.}= \sqrt{2E_{\rm
    kin}/\mu_{\rm a}} = \hbar k_n/\mu_a$ is simply
\begin{equation}
   {\sf K}_{n'\leftarrow n}(E) =
   \frac{\pi \hbar}{\mu_a k_n}
    \sum_{\ell=0}^\infty  \sum_{J=|\ell-j|}^{\ell +j}
	\left(\frac{2J+1}{2j+1}\right) P^J_{n,\ell}(E)
    \;\; \frac{\sum_{\ell'} P^J_{n',\ell'}(E) }
    {\sum_{n'',\ell''}P^J_{n'',\ell''}(E)},
\end{equation}
where $\ell' = |J-j'|, \dots, J+j' $ and $\ell'' = |J-j''|, \dots,
J+j'' $, respectively. As in the case of Eqs.(\ref{eq:K-Q}) and
(\ref{eq:K-R}), we obtain the total quenching and reaction rate
constant by summing over the appropriate channels and arrangements
\begin{eqnarray}
 {\sf K}_n^{\rm Q}(E) & = & \frac{\pi \hbar}{\mu_a k_n}
    \sum_{\ell=0}^\infty \sum_{J=|\ell-j|}^{\ell +j}
	\left(\frac{2J+1}{2j+1}\right) P^J_{n,\ell}(E)  \frac{P^J_Q}{P^J_{\rm tot}}
\label{eq:K-Q-stat} , \\
 {\sf K}_n^{\rm R}(E) & = &  \frac{\pi \hbar}{\mu_a k_n}
    \sum_{\ell=0}^\infty  \sum_{J=|\ell-j|}^{\ell +j}
	\left(\frac{2J+1}{2j+1}\right) P^J_{n,\ell}(E) \frac{P^J_R}{P^J_{\rm tot}}
	\label{eq:K-R-stat} ,
\end{eqnarray}
where
\begin{equation}
\fl   P^J_Q (E) \equiv \sum_{\scriptsize \begin{array}{c}n'\neq n\\ a'=a \end{array}}
   \sum_{\ell'}  P^J_{n',\ell'}(E), \qquad  
   P^J_R (E) \equiv  \sum_{\scriptsize \begin{array}{c}n'\neq n\\ a'\neq a \end{array}}
   \sum_{\ell'}  P^J_{n',\ell'}(E) \; ,   
\end{equation}
with $P^J_{\rm tot}(E) =   P^J_Q (E) +  P^J_R (E) + \sum_{\ell ''} P^J_{n,\ell''}(E)$.

Due to the small mass of our system, we use the PST capture model to
determine $P^J_{n',\ell'}(E)$ in the product channels only, where the
kinetic energy is relatively large and the effect of near-threshold
states below dynamical barriers may be neglected \cite{dulieu}, i.e.,
\begin{equation}
\fl    P^J_{n',\ell'}(E) = \left\{\begin{array}{ll} 1, & E-E_{n'} \geq V_{b,n'} \\
      0, & \mbox{otherwise} \end{array}\right.  \qquad
      V_{b,n'} = \sqrt{\frac{[\hbar^2 \ell'(\ell' +1)]^3}{54\mu_{a'}^3 C_{6,n'}}} .
\end{equation}
The van der Waals coefficient $C_{6,n'}$ and reduced mass $\mu_{a'}$
appearing in the expression of the centrifugal barrier height
$V_{b,n'}$ depends in the particular channel $n'=(a',v',j')$.  For
$P^J_{n,\ell}(E)$ in the entrance channel, we use the low-energy
expression
\begin{equation}
  P^J_{n,\ell}(E) = A^J_{n,\ell}(E) k_n^{2\ell +1} \; ,
\end{equation}
where the coefficient $A^J_{n,\ell}(E)$ is related to Jost functions
\cite{taylor}.

\begin{figure}[t]
\includegraphics[clip,width= 0.99\textwidth]{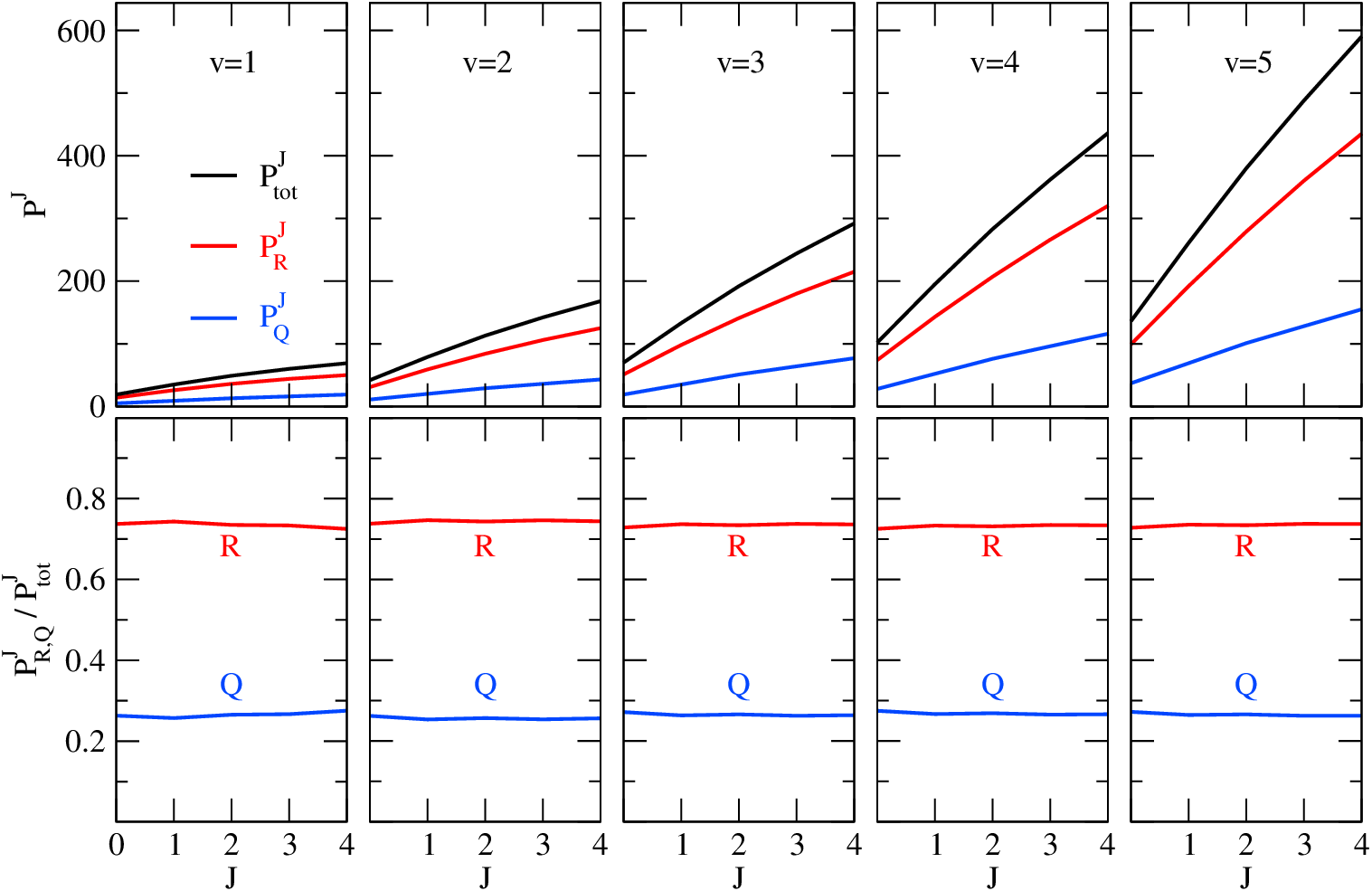}
\caption{Top panels: $P^J_{Q(R)}$ and $P^J_{\rm tot}$ for para-H$_2$
  as a function of $J$ for different entrance channels $n=(j=0,v)$
  with $v=1, \dots , 5$ ($v=0$ is not shown since only
  reaction is allowed, so that $P^J_R/P^J_{\rm
    tot}=1$). Lower panels: corresponding ratios $P^J_{Q(R)}/P^J_{\rm
    tot}$. Also shown is the number of molecular states opened to
  quenching and reaction, $N_Q$, and $N_R$ respectively, and their
  ratios.  These are $J$-independent and equal to the $P^{J=0}_{Q(R)}$
  values.  }
\label{fig:R/Q-para}
\end{figure}

\begin{figure}[t]
\includegraphics[clip,width= 0.99\textwidth]{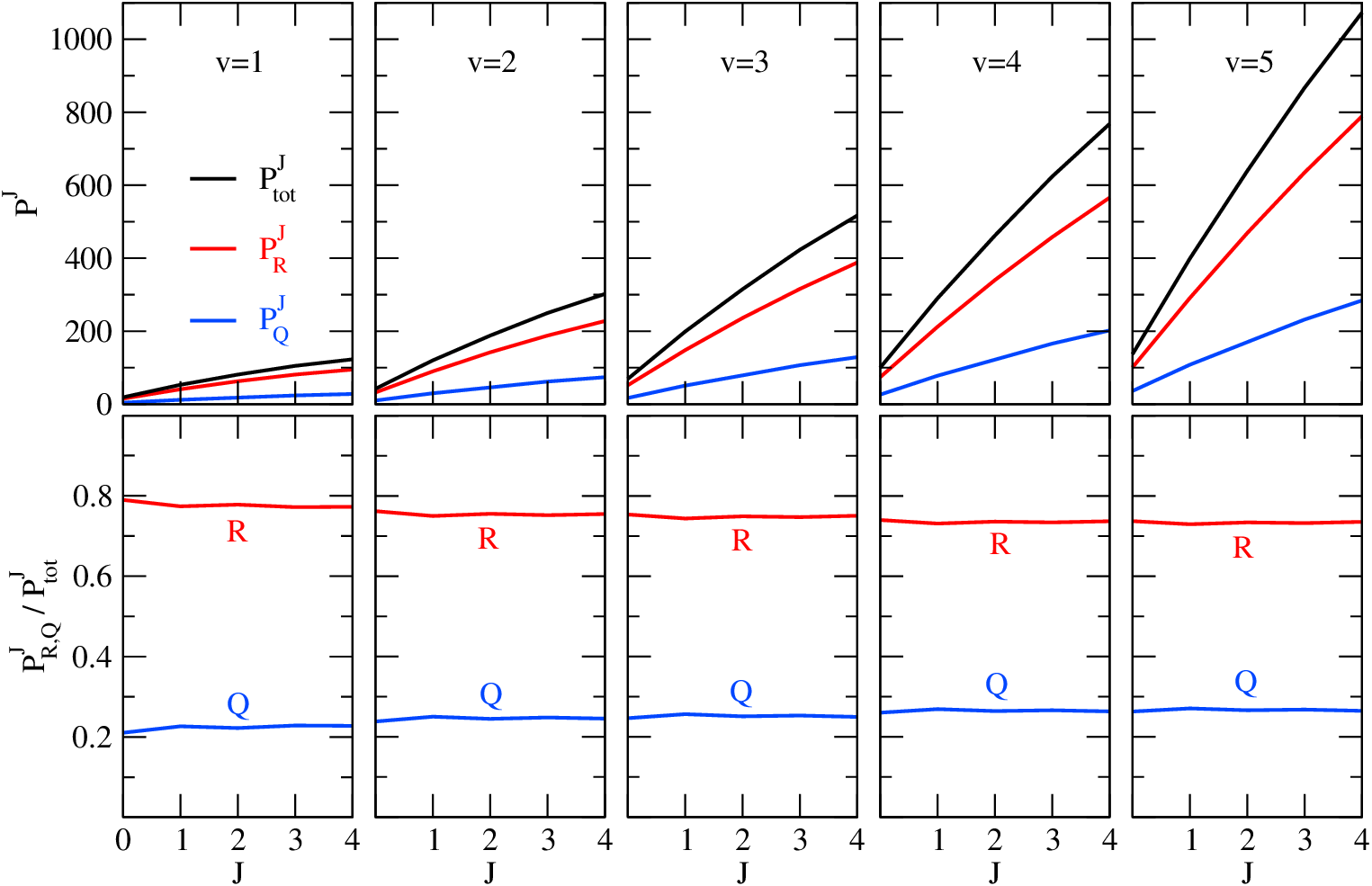}
\caption{Same as Fig.~\ref{fig:R/Q-para}, but for ortho-H$_2$
  initially in $j=1$.  }
\label{fig:R/Q-ortho}
\end{figure}

\begin{figure}[t]
\includegraphics[clip,width= 0.99\textwidth]{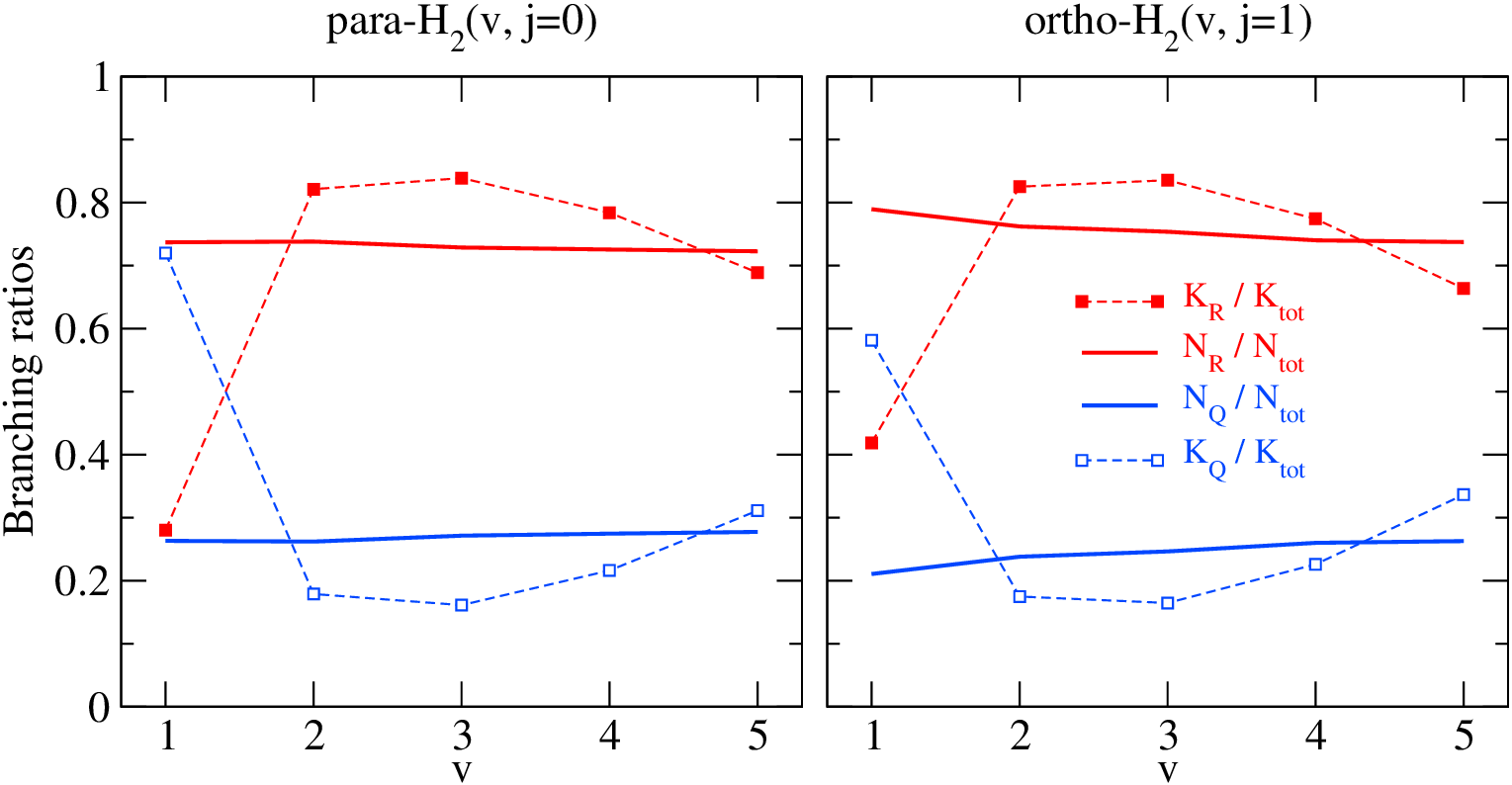}
\caption{Comparison of the branching ratios of quenching and reaction
  processes between the full quantum treatment $K_n^{Q(R)}/K^{\rm
    tot}_n$ (dash lines) and the statistical model $N^{Q(R)}/N^{\rm
    tot}$ (solid lines) for para-H$_2$ and ortho-H$_2$. Results for
  $v=0$ are not shown, since only reactive scattering is
  allowed. Ratios agree within 10--20\%, except for $v=1$
    for which the effect of the reaction barrier is more important.}
\label{fig:comparison}
\end{figure}

At low-energy, $\sum_{\ell ''} P^J_{n,\ell''}(E)$ is small and we can
omit it in evaluating $P^J_{\rm tot}(E) \simeq P^J_Q (E) + P^J_R (E)$.
Furthermore, $P^J_Q$, $P^J_R$, and $P^J_{\rm tot}$ are essentially
$E$-independent at low-energy; figures~\ref{fig:R/Q-para} and
\ref{fig:R/Q-ortho} show their variation with $J$ for the initial
state $n=(v,j=0)$ for para-H$_2$ and $n=(v, j=1)$ for ortho-H$_2$,
respectively. Although $P^J_Q$, $P^J_R$, and $P^J_{\rm tot}$ all
increase with $J$, the ratios $P^J_Q/P^J_{\rm tot}$ and
$P^J_R/P^J_{\rm tot}$ are nearly constant (roughly 73\% reaction and
27\% quenching), allowing us to further simplify
eq~(\ref{eq:K-Q-stat}) and (\ref{eq:K-R-stat}) as
\begin{equation}
 {\sf K}_n^{\rm Q(R)}(E) = \frac{\pi \hbar}{\mu_a}  \frac{P_{Q(R)}}{P_{\rm tot}}
    \sum_{\ell=0}^\infty k_n^{2\ell} \sum_{J=|\ell-j|}^{\ell +j}
	\left(\frac{2J+1}{2j+1}\right) A^J_{n,\ell}(E) 
	 , \label{eq:K-Q/R-stat} 
\end{equation}
where $P^J_{Q(R)}/P^J_{\rm tot} \simeq P_{Q(R)}/P_{\rm tot}$ does not
depend on $J$. In particular, their values for $J=0$ is simply given
by the corresponding number of open channels $N_Q$ for quenching and
$N_R$ for reaction, so that $P_{Q(R)}/P_{\rm tot}\simeq
P^{J=0}_{Q(R)}/P^{J=0}_{\rm tot} = N_{Q(R)}/N_{\rm tot}$ with $N_{\rm
  tot} =N_Q+N_R$.  Although the rate constants for quenching and
reaction processes depends on the details of the entrance channel via
the sums over $\ell$ and $J$, their ratio will not; the double-sum
cancels out, and we find
\begin{equation}
  \frac{{\sf K}_n^{\rm R}(E)}{{\sf K}_n^{\rm Q}(E)} 
  \simeq \frac{P_R}{P_Q} \simeq  \frac{N_R}{N_Q}
,\qquad
   \frac{{\sf K}_n^{\rm Q(R)}(E)}{{\sf K}_n^{\rm tot}(E)} 
   \simeq \frac{P_{Q(R)}}{P_{\rm tot}} 
   \simeq  \frac{N_{Q(R)}}{N_{\rm tot}}.
\end{equation}
Figure~\ref{fig:comparison} compares the results obtained using this
statistical model and the full quantum treatment, We find that the
agreement is roughly within 10--20\% for $v\geq2$, but quickly becomes
worse when the reaction barrier plays an important role for very low
$v$. Overall, this statistical model based on PST approximately gives
the right branching ratio between quenching and reaction processes.

As discussed in previous sections, the magnitude of the rate constants
at low scattering energies are mainly given by the $s$-wave
contribution: figures~\ref{fig:sigma-j-2by6-para} and
\ref{fig:sigma-j-2by6-ortho} show that this is the case for most of
initial $v$ up to 10~mK and higher.  For para-H$_2$ in $j=0$, the
$s$-wave $\ell =0$ implies $J=0$ only, so that $(2J+1)/(2j+1) = 1$,
while for ortho-H$_2$ in $j=1$, $\ell =0$ implies $J=1$ only, so that
$(2J+1)/(2j+1) = 1$ as well.  In both cases, both sums over $\ell$ and
$J$ in Eq.~(\ref{eq:K-Q/R-stat}) reduce to a single term;
\begin{equation} 
     {\sf K}_n^{\rm Q(R)}(E) = \frac{\pi\hbar}{\mu_a}\frac{P_{Q(R)}}{P_{\rm tot}}
    \times \left\{ \begin{array}{ll}
     A^{J=0}_{n,\ell=0}(E), & \mbox{ for para-H$_2$ in $j=0$} \\ & \\
     A^{J=1}_{n,\ell=0}(E), & \mbox{ for ortho-H$_2$ in $j=1$} \end{array}\right.
	 . \label{eq:K-Q/R-stat-para/ortho} 
\end{equation}
For $\ell=0$ with $J=j$, the coefficient
$A^{J=j}_{n,\ell=0}(E\rightarrow 0)$ is simply related to the
imaginary component $\beta_{v,j}$ of the complex scattering length
$a_{v,j} = \alpha_{v,j} - i\beta_{v,j}$ \cite{bala-alpha-beta} in the
entrance channel $n=(a,v,j)$ of the initial arrangement $a$=H$_2$+D
and total angular momentum $J=j$ \cite{NTR},
\begin{equation}
   A^{J=j}_{n,\ell=0}(E\rightarrow 0) = 4 \beta_{v,j} \;,
\end{equation}
such that the rate constant takes the simple form
\begin{equation} 
 \fl    {\sf K}_n^{\rm Q/R}(E) = \frac{4\pi \hbar}{\mu_a}  
     \frac{P_{Q/R}}{P_{\rm tot}} \beta_{v,j},
      \qquad\mbox{ with }
    \left\{ \begin{array}{ll}
     J=j=0, & \mbox{ for para-H$_2$} \\ & \\
     J=j=1, & \mbox{ for ortho-H$_2$} \end{array}\right.
	 . \label{eq:K-Q/R-stat-para/ortho-s-wave} 
\end{equation}
Table~\ref{tab1} contains the values of $\alpha_{v,j}$ and
$\beta_{v,j}$, which were extracted from the full quantum calculations
\cite{note-scattering-length}.  They can be compared to an approximate
value based on quantum reflection
\cite{julienne-universal,cote-heller,cote-friedrich}, which is often
used for barrierless reactions,
\begin{equation}
   \beta_{v,j} \simeq \frac{4\pi}{\Gamma ( 1/4 )^2}  
   \left( \frac{2\mu_a C^{v,j}_6}{\hbar^2} \right)^{1/4}.
   \label{eq:beta-approx}
\end{equation}
Using the value
$C_6^{v=0,j=0}=7.053$~a.u.~\cite{PRL-rydberg-dressing}, together with
$\mu_a \simeq 1836$ a.u., we obtain $\beta \simeq 6.06$ a.u., which is
many order of magnitude larger than the values of $\beta$ from the
full quantum treatment. This illustrates well the fact that, although
the statistical model gives approximately the right branching ratio of
quenching to reaction processes, the approximation in
Eq.~(\ref{eq:beta-approx}) cannot be used to obtain the magnitude of
those processes. Although such an approximation can be applied to
barrierless chemical systems, the reaction barrier reduces the value
of $\beta$ considerably, as shown in Table~\ref{tab1}.

\begin{table}[ht]
\caption{\label{tab1}Real and imaginary contributions to the
  scattering length $a_{v,j}=\alpha_{v,j}-i\beta_{v,j}$ for para-H$_2$
  ($j=0$) and ortho-H$_2$ ($j=1$) extracted from the quantum results.}
\begin{indented}
\lineup
\item[]\begin{tabular}{*{5}{l}}
\br
      &    \centre{2}{D + para-H$_2(v,j=0)$}    &    \centre{2}{D +
ortho-H$_2(v,j=1)$} \\
\ns
      &     \crule{2}          &     \crule{2}        \\
\ns
 $v$  &  $\alpha$ (a.u.)  &  $\beta$ (a.u.)  &  $\alpha$ (a.u.)  &  $\beta$
(a.u.)\\
\mr
  0   &   14.8     &  $7.9\times 10^{-14}$   &   14.7   & $2.8\times 10^{-13}$\\
  1   &   14.0     &  $2.4\times 10^{-7}$    &   13.9   & $3.1\times 10^{-7}$\\
  2   &   13.4     &  $7.2\times 10^{-5}$    &   13.2   & $1.1\times 10^{-4}$\\
  3   &   12.9     &  $1.2\times 10^{-3}$    &   12.7   & $2.5\times 10^{-3}$\\
  4   &   12.2     &  $2.4\times 10^{-2}$    &   11.9   & $7.3\times 10^{-2}$\\
  5   &  \07.0     &  $3.0\times 10^{-1}$    &  \06.6   & $6.6\times 10^{-1}$\\
\br
\end{tabular}
\end{indented}
\end{table}

\section{Conclusion}
\label{sec:conclusion} 

The nuclear spin symmetry restricts the possible rotational states of
H$_2$, leading to para-H$_2$ with even $j$ and ortho-H$_2$ with odd
$j$. We investigated the effect of these restrictions on the
scattering of H$_2$ with D, which can lead to elastic collisions, to
quenching H$_2$, or reaction to produce HD. We computed these various
processes for the six lowest vibrational levels $v$ of H$_2$. We found
structures in the energy range corresponding to 100 mK and 10 K, due
to low partial waves. Except for cases involving a sharp $p$-wave
resonance, the results for para- and ortho-H$_2$ are very similar,
with structures due to $d$ or $f$-waves located at roughly the same
energies in both cases, and with rate constants of comparable
magnitude.

There are marked differences for the cases in which a $p$-wave
resonance is present, namely the three lowest $v$ levels. For example,
in the reactive case, while the sharp peak is present for both para-
and ortho-H$_2$ in $v=0$, it is absent for ortho-H$_2$ in $v=1$ and
para-H$_2$ in $v=2$. The same is true for the quenching rate constant
(though there is no quenching for $v=0$, since it is the ground
state). We also find the same overall behavior for the elastic cross
sections. We also note that the exact position of the resonance is
slightly affected by the initial value of $j$.

We also investigated the state-to-state product formation, and found
that the branching ratios are not sensitive to the scattering energy
in the range considered here for para-H$_2$. The same is true for
ortho-H$_2$ except for the lower initial levels ($v=1,\ 2$), where the
$p$-wave resonance affects the branching ratios. By examining the
rotationally-resolved rate constants of a specific initial case,
namely $v=2$, we can trace this dependence to sharper increases of the
inelastic (either quenching or reaction) rate constant for certain
rotational products near the resonance. These variations are due to
the different coupling strengths of each total angular momentum $J=0$,
1, and 2 contributing to $p$-wave ($\ell=1$) initial partial wave for
ortho-H$_2$, as opposed to the case of para-H$_2$, when only $J=1$
contributes.

The range of energies where the structure is present should be
reachable experimentally, and thus could probe the effect of the
nuclear symmetry on reaction (or quenching). In addition, since the
occurrence of $p$-wave resonances is very sensitive to the exact shape
of the potential energy surface, detecting any structure in this
benchmark system will provide a unique tool to compare with {\it ab
  initio} quantum chemistry calculations.  Finally, based on our
results, we suspect that the effect of the nuclear spin symmetry on
cold and ultracold chemistry will be significant in systems displaying
resonant features, and less so otherwise. However, if the nuclear spin
is involved in the dynamics of the system, e.g., through an
interaction term in the Hamiltonian, as in the studies of
Refs.~\cite{krems-PRA-2007,Hutson-PRA-2011,
  Halvick-PCCP-2011,Launay-Laser-2006,Lique-Astro-2012,Lique-JCP-2014},
the nuclear spin symmetry could play an even more important role.
Finally, we compared the results of our full quantum treatment to a
statistical model, and found that although the model reproduces the
proportion of quenching and reaction processes within 20\% in most
cases, it fails to give the magnitude of the rate constants.

\section*{Acknowledgments}

The authors wish to thank Prof. Suhbas Ghosal for fruitful
discussions.  This work was partially supported by supported by the
U.S. Department of Energy, Office of Basic Energy Science (R.C.),
and the Army Research Office Chemistry Division (I. S.).

\section*{References}

\bibliographystyle{unsrt}
\bibliography{simbotin_ref}

\begin{thebibliography}{10}

\bibitem{cote-1997}
R~C\^ot\'e and A~Dalgarno.
\newblock {Mechanism for the production of vibrationally excited ultracold
  molecules of $^7$Li$_2$}.
\newblock {\em {Chem. Phys. Lett.}}, {279}({1-2}):{50--54}, {NOV 7} {1997}.

\bibitem{jmp-1999}
R.~C\^ot\'e and A.~Dalgarno.
\newblock Mechanism for the production of $^6${Li}$_2$ and $^7${Li}$_2$
  ultracold molecules.
\newblock {\em J. Mol. Spectr.}, 195(2):236 -- 245, 1999.

\bibitem{pillet-1998}
A~Fioretti, D~Comparat, A~Crubellier, O~Dulieu, F~Masnou-Seeuws, and P~Pillet.
\newblock Formation of cold {Cs}$_2$ molecules through photoassociation.
\newblock {\em Phys. Rev. Lett.}, 80(20):4402--4405, MAY 18 1998.

\bibitem{knize-1998}
T~Takekoshi, B~M Patterson, and R~J Knize.
\newblock {Observation of optically trapped cold cesium molecules}.
\newblock {\em {Phys. Rev. Lett.}}, {81}({23}):{5105--5108}, {DEC 7} {1998}.

\bibitem{gospel:09}
R.V. Krems, W.C. Stwalley, and B.~Friedrich.
\newblock {\em Cold molecules: theory, experiment, applications}.
\newblock CRC Press, 2009.

\bibitem{ian.smith:08}
I.W.M. Smith.
\newblock {\em Low temperatures and cold molecules}.
\newblock Imperial College Press, 2008.

\bibitem{w+z:09}
M.~Weidem\"uller and C.~Zimmermann.
\newblock {\em Cold atoms and molecules: a testground for fundamental many
  particle physics}.
\newblock Physics textbook. Wiley-VCH, 2009.

\bibitem{stwalley:canjchem:04}
W.~C. Stwalley.
\newblock Collisions and reactions of ultracold molecules.
\newblock {\em Can.~J.~Chem.}, 82(6):709--712, 2004.

\bibitem{weck+bala:irpc:06}
Philippe~F. Weck and N.~Balakrishnan.
\newblock {Importance of long-range interactions in chemical reactions at cold
  and ultracold temperatures}.
\newblock {\em Int. Rev. Phys. Chem.}, {25}({3}):{283--311}, {JUL-SEP} {2006}.

\bibitem{jeremy:irpc:06}
Pavel Sold\'an and Jeremy~M. Hutson.
\newblock Molecule formation in ultracold atomic gases.
\newblock {\em Int. Rev. Phys. Chem.}, 25:497--526, 2006.

\bibitem{jeremy:irpc:07}
Pavel Sold\'an and Jeremy~M. Hutson.
\newblock Molecular collisions in ultracold atomic gases.
\newblock {\em Int. Rev. Phys. Chem.}, 26:1--28, 2007.

\bibitem{krems:pccp:08}
R.~V. Krems.
\newblock Cold controlled chemistry.
\newblock {\em Phys. Chem. Chem. Phys.}, 10:4079--4092, 2008.

\bibitem{FOPA+STIRAP}
Elena Kuznetsova, Marko Gacesa, Philippe Pellegrini, Susanne~F Yelin, and Robin
  C\^ot\'e.
\newblock Efficient formation of ground-state ultracold molecules via {STIRAP}
  from the continuum at a {Feshbach} resonance.
\newblock {\em New J. Phys.}, 11(5):055028, 2009.

\bibitem{Thermo-K2Rb2}
Jason~N. Byrd, John~A. Montgomery, and Robin C\^ot\'e.
\newblock Structure and thermochemistry of {K}$_{2}${Rb}, {KRb}$_{2}$, and
  {K}$_{2}${Rb}${}_{2}$.
\newblock {\em Phys. Rev. A}, 82:010502, Jul 2010.

\bibitem{junye:sci:10}
S.~Ospelkaus, K.-K. Ni, D.~Wang, M.~H.~G. de~Miranda, B.~Neyenhuis,
  G.~Qu\'em\'ener, P.~S. Julienne, J.~L. Bohn, D.~S. Jin, and J.~Ye.
\newblock Quantum-state controlled chemical reactions of ultracold
  potassium-rubidium molecules.
\newblock {\em Science}, 327:853, 2010.

\bibitem{zare-private}
Richard~N. Zare.
\newblock private communication, 2014.

\bibitem{zare-PNAS}
Justin Jankunas, Mahima Sneha, Richard~N. Zare, Foudhil Bouakline, Stuart~C.
  Althorpe, Diego Herr\'aez-Aguilar, and F.~Javier Aoiz.
\newblock Is the simplest chemical reaction really so simple?
\newblock {\em Proc. Natl. Acad. Sci.}, 111(1):15--20, 2014.

\bibitem{astro-cloud}
Daniele Galli and Francesco Palla.
\newblock Deuterium chemistry in the primordial gas.
\newblock {\em Planet. Space Sci.}, 50(12–13):1197 -- 1204, 2002.
\newblock Special issue on Deuterium in the Universe.

\bibitem{astro}
C.~D. Gay, P.~C. Stancil, S.~Lepp, and A.~Dalgarno.
\newblock The highly deuterated chemistry of the early universe.
\newblock {\em Astrophys. J.}, 737(1):44, 2011.

\bibitem{Schatz:JCP:1987}
Toshiyuki Takayanagi, Nobuyuki Masaki, Kazutaka Nakamura, Makoto Okamoto, Shin
  Sato, and George~C. Schatz.
\newblock The rate constants for the {H + H}$_2$ reaction and its isotopic
  analogs at low temperatures: Wigner threshold law behavior.
\newblock {\em J. Chem. Phys.}, 86(11):6133--6139, 1987.

\bibitem{bala:fh2:cpl01}
N~Balakrishnan and A~Dalgarno.
\newblock {Chemistry at ultracold temperatures}.
\newblock {\em {Chem. Phys. Lett.}}, {341}({5-6}):{652--656}, {JUN 29} {2001}.

\bibitem{bodo:fd2:jpb02}
E~Bodo, FA~Gianturco, and A~Dalgarno.
\newblock {The reaction of F+D$_2$ at ultra-low temperatures: the effect of
  rotational excitation}.
\newblock {\em {J. Phys. B - Atom. Mol. Op. Phys.}}, {35}({10}):{2391--2396},
  {MAY 28} {2002}.

\bibitem{Bala:PRL:Heh2}
N.~Balakrishnan, R.~C. Forrey, and A.~Dalgarno.
\newblock Quenching of {H}$_{2}$ vibrations in ultracold $^{3}${He} and
  $^{4}${He} collisions.
\newblock {\em Phys. Rev. Lett.}, 80(15):3224--3227, Apr 1998.

\bibitem{bala:KrArH2:pra09}
N.~Balakrishnan, Bradley~C. Hubartt, Luke Ohlinger, and Robert~C. Forrey.
\newblock {Noble-gas quenching of rovibrationally excited H$_2$}.
\newblock {\em {Phys. Rev. A}}, {80}({1}):{012704}, {JUL} {2009}.

\bibitem{bala-PRA2008}
Goulven Qu\'em\'ener, Naduvalath Balakrishnan, and Roman~V. Krems.
\newblock Vibrational energy transfer in ultracold molecule-molecule
  collisions.
\newblock {\em Phys. Rev. A}, 77:030704, Mar 2008.

\bibitem{bala-JCP2009}
Goulven Qu\'em\'ener and Naduvalath Balakrishnan.
\newblock Quantum calculations of {H$_2$-H$_2$} collisions: From ultracold to
  thermal energies.
\newblock {\em J. Chem. Phys.}, 130(11):114303, 2009.

\bibitem{bala-JCP2011}
S.~Fonseca~dos Santos, N.~Balakrishnan, S.~Lepp, G.~Qu\'m\'ener, R.~C. Forrey,
  R.~J. Hinde, and P.~C. Stancil.
\newblock Quantum dynamics of rovibrational transitions in {H$_2$-H$_2$}
  collisions: Internal energy and rotational angular momentum conservation
  effects.
\newblock {\em J. Chem. Phys.}, 134(21):214303, 2011.

\bibitem{bohn-PRA2001}
Alexandr~V. Avdeenkov and John~L. Bohn.
\newblock Ultracold collisions of oxygen molecules.
\newblock {\em Phys. Rev. A}, 64:052703, Oct 2001.

\bibitem{bohn-JCP2003}
Alessandro Volpi and John~L. Bohn.
\newblock Fine-structure effects in vibrational relaxation at ultralow
  temperatures.
\newblock {\em J. Chem. Phys.}, 119(2):866--871, 2003.

\bibitem{abc:cpc:2k}
D~Skouteris, JF~Castillo, and DE~Manolopoulos.
\newblock {ABC: a quantum reactive scattering program}.
\newblock {\em {Comp. Phys. Comm.}}, {133}({1}):{128--135}, {DEC 1} {2000}.

\bibitem{Delves:58}
L.~M. Delves.
\newblock Tertiary and general-order collisions.
\newblock {\em Nuc. Phys.}, 9(3):391--399, 1958-1959.

\bibitem{manolopoulos:logder:jcp86}
D.~E. Manolopoulos.
\newblock An improved log derivative method for inelastic scattering.
\newblock {\em J. Chem. Phys.}, 85(11):6425--6429, 1986.

\bibitem{miranda:hd2:jcp98}
Marcelo~P. de~Miranda, David~C. Clary, Jesus~F. Castillo, and David~E.
  Manolopoulos.
\newblock Using quantum rotational polarization moments to describe the
  stereodynamics of the {H + D}$_2(v = 0,\ j = 0)\rightarrow$ {HD}$(v',\ j')$ +
  {D} reaction.
\newblock {\em J. Chem. Phys.}, 108(8):3142--3153, 1998.

\bibitem{jesus:hd2:pccp10}
Jesus Aldegunde, P.~G. Jambrina, Vicente Saez-Rabanos, Marcelo~P. de~Miranda,
  and F.~J. Aoiz.
\newblock {Quantum mechanical mechanisms of inelastic and reactive H +
  D$_2(v=0,\ j=2)$ collisions}.
\newblock {\em {Phys. Chem. Chem. Phys.}}, {12}({41}):{13626--13636}, {2010}.

\bibitem{castillo:fh2:jcp96}
Jesus~F. Castillo, David~E. Manolopoulos, Klaus Stark, and Hans-Joachim Werner.
\newblock Quantum mechanical angular distributions for the {F+H}$_2$ reaction.
\newblock {\em J. Chem. Phys.}, 104(17):6531--6546, 1996.

\bibitem{mano:fhd:fdiscuss98}
JF~Castillo and DE~Manolopoulos.
\newblock Quantum mechanical angular distributions for the {F+HD} reaction.
\newblock {\em Faraday Discuss. Chem. Soc.}, 110:119--138, 1998.

\bibitem{skouteris:clhd:science99}
D~Skouteris, DE~Manolopoulos, WS~Bian, HJ~Werner, LH~Lai, and KP~Liu.
\newblock {van der Waals interactions in the Cl+HD reaction}.
\newblock {\em {Science}}, {286}({5445}):{1713--1716}, {NOV 26} {1999}.

\bibitem{bala:clhd:JCP04}
N~Balakrishnan.
\newblock {On the role of van der Waals interaction in chemical reactions at
  low temperatures}.
\newblock {\em {J. Chem. Phys.}}, {121}({12}):{5563--5566}, {SEP 22} {2004}.

\bibitem{bala:fhcl:jcp08}
Goulven Qu\'em\'ener and Naduvalath Balakrishnan.
\newblock {Cold and ultracold chemical reactions of F+HCl and F+DCl}.
\newblock {\em {J. Chem. Phys.}}, {128}({22}):{224304}, {JUN 14} {2008}.

\bibitem{jesus:fh2:jcp06}
J.~Aldegunde, J.~M. Alvarino, M.~P. de~Miranda, V.~Saez Rabanos, and F.~J.
  Aoiz.
\newblock {Mechanism and control of the F+H$_2$ reaction at low and ultralow
  collision energies}.
\newblock {\em {J. Chem. Phys.}}, {125}({13}):{133104}, {OCT 7} {2006}.

\bibitem{PCCP-H2+D}
Ion Simbotin, Subhas Ghosal, and Robin C\^ot\'e.
\newblock A case study in ultracold reactive scattering: {D + H}$_2$.
\newblock {\em Phys. Chem. Chem. Phys.}, 13:19148--19155, 2011.

\bibitem{mielke}
Steven~L. Mielke, Kirk~A. Peterson, David~W. Schwenke, Bruce~C. Garrett,
  Donald~G. Truhlar, Joe~V. Michael, Meng-Chih Su, and James~W. Sutherland.
\newblock H + {H}$_{2}$ thermal reaction: A convergence of theory and
  experiment.
\newblock {\em Phys. Rev. Lett.}, 91:063201, Aug 2003.

\bibitem{bkmp2:jcp96}
AI~Boothroyd, WJ~Keogh, PG~Martin, and MR~Peterson.
\newblock {A refined H$_3$ potential energy surface}.
\newblock {\em J. Chem. Phys.}, {104}({18}):{7139--7152}, {MAY 8} {1996}.

\bibitem{PRL-rydberg-dressing}
Jia Wang, Jason~N. Byrd, Ion Simbotin, and R.~C\^ot\'e.
\newblock Tuning ultracold chemical reactions via {Rydberg}-dressed
  interactions.
\newblock {\em Phys. Rev. Lett.}, 113:025302, Jul 2014.

\bibitem{krems-PRA-2007}
T.~V. Tscherbul, J.~K\l{}os, L.~Rajchel, and R.~V. Krems.
\newblock Fine and hyperfine interactions in cold {YbF-He} collisions in
  electromagnetic fields.
\newblock {\em Phys. Rev. A}, 75:033416, Mar 2007.

\bibitem{Hutson-PRA-2011}
Maykel~L. Gonz\'alez-Mart\'{\i}nez and Jeremy~M. Hutson.
\newblock Effect of hyperfine interactions on ultracold molecular collisions:
  {NH$(^3\Sigma^-)$} with {Mg$(^1S)$} in magnetic fields.
\newblock {\em Phys. Rev. A}, 84:052706, Nov 2011.

\bibitem{Halvick-PCCP-2011}
T.~Stoecklin and Ph. Halvick.
\newblock Collisional relaxation of {MnH$(X^7\Sigma^+)$} in a magnetic field:
  effect of the nuclear spin of {Mn}.
\newblock {\em Phys. Chem. Chem. Phys.}, 13:19142--19147, 2011.

\bibitem{Launay-Laser-2006}
A.~Simoni and J.-M. Launay.
\newblock Ultracold atom-molecule collisions with hyperfine coupling.
\newblock {\em Laser Phys.}, 16(4):707--712, 2006.

\bibitem{Lique-Astro-2012}
A.~Faure and F.~Lique.
\newblock The impact of collisional rate coefficients on molecular hyperfine
  selective excitation.
\newblock {\em Month. Not. Roy. Astro. Soc.}, 425(1):740--748, 2012.

\bibitem{Lique-JCP-2014}
M.~Lanza and F.~Lique.
\newblock Hyperfine excitation of linear molecules by para- and ortho-{H}$_2$:
  Application to the {HCl--H}$_2$ system.
\newblock {\em J. Chem. Phys.}, 141(16):164321, 2014.

\bibitem{miller}
William~H. Miller.
\newblock Study of the statistical model for molecular collisions.
\newblock {\em J. Chem. Phys.}, 52(2):543--551, 1970.

\bibitem{dulieu}
Maykel~L. Gonz\'alez-Mart\'\i{nez}, Olivier Dulieu, Pascal Larr\'egaray, and
  Laurent Bonnet.
\newblock Statistical product distributions for ultracold reactions in external
  fields.
\newblock {\em Phys. Rev. A}, 90:052716, Nov 2014.

\bibitem{taylor}
J.R. Taylor.
\newblock {\em Scattering Theory}.
\newblock Dover Books on Engineering. Dover Publications, 2012.

\bibitem{bala-alpha-beta}
N.~{Balakrishnan}, V.~{Kharchenko}, R.~C. {Forrey}, and A.~{Dalgarno}.
\newblock {Complex scattering lengths in multi-channel atom-molecule
  collisions}.
\newblock {\em Chem. Phys. Lett.}, 280:5--9, November 1997.

\bibitem{NTR}
I.~Simbotin, S.~Ghosal, and R.~C\^ot\'e.
\newblock Threshold resonance effects in reactive processes.
\newblock {\em Phys. Rev. A}, 89:040701, Apr 2014.

\bibitem{note-scattering-length}
The values of $\alpha$ and $\beta$ are comparable to those obtained in
  \cite{PCCP-H2+D}, except for $v=5$ because in this work we used a truncation
  energy $E_{\max}=3.5$ {eV}, which is lower than the value used in
  \cite{PCCP-H2+D}.

\bibitem{julienne-universal}
Zbigniew Idziaszek and Paul~S. Julienne.
\newblock Universal rate constants for reactive collisions of ultracold
  molecules.
\newblock {\em Phys. Rev. Lett.}, 104:113202, Mar 2010.

\bibitem{cote-heller}
R.~C\^ot\'e, E.~J. Heller, and A.~Dalgarno.
\newblock Quantum suppression of cold atom collisions.
\newblock {\em Phys. Rev. A}, 53:234--241, Jan 1996.

\bibitem{cote-friedrich}
R.~C\^ot\'e, H.~Friedrich, and J.~Trost.
\newblock Reflection above potential steps.
\newblock {\em Phys. Rev. A}, 56:1781--1787, Sep 1997.

\end{thebibliography}

\end{document}